\documentclass[aps,prd,preprint,tightenlines,preprintnumbers,showpacs,superscriptaddress,nofootinbib,eqsecnum,floatfix]{revtex4-1}

\usepackage{bm}
\usepackage{graphicx}
\usepackage[pdftex]{color}
\usepackage{amsmath, amssymb}
\usepackage{hyperref}
\usepackage{dcolumn}
\usepackage{array}
\hypersetup{
    colorlinks=true,     % false: boxed links; true: colored links
    linkcolor=blue,      % color of internal links
    citecolor=blue,      % color of links to bibliography
    filecolor=blue,      % color of file links
    urlcolor=blue        % color of external links
}
\usepackage{rotating}
\usepackage{mathtools}
\usepackage{multirow}
\usepackage{slashed} % For Feynman slashed notation

\definecolor{red}{rgb}{0.8,0.0,0.0}
\definecolor{green}{rgb}{0.0,0.6,0.0}
\definecolor{darkblue}{rgb}{0.0,0.1,0.7}
\definecolor{brown}{rgb}{0.6,0.1,0.0}
\definecolor{gray}{rgb}{0.6,0.6,0.6}
\definecolor{darkgreen}{rgb}{0.0, 0.545098, 0.0}
\definecolor{purple}{rgb}{0.5,0.0,0.5}
\definecolor{applegreen}{rgb}{0.55, 0.71, 0.0}
\definecolor{babypink} {rgb}{0.64, 0.44, 0.44}
\definecolor{orange}{rgb}{0.9,0.4,0.0}

\newcommand{\eg}{\emph{e.g.}}

\newcommand{\bi}{\begin{itemize}}
\newcommand{\ei}{\end{itemize}}

\newcommand{\be}{\begin{equation}}
\newcommand{\ee}{\end{equation}}

\newcommand{\bea}{\begin{eqnarray}}
\newcommand{\eea}{\end{eqnarray}}

\newcolumntype{P}[1]{>{\centering\arraybackslash}p{#1}} %To center "paragraph" columns in tables

\newcommand{\ket}[1] {|#1\rangle}

\newcommand{\Dslash}{\ensuremath{D\kern-0.6em/\kern0.15em}}
\newcommand{\tev}{\ensuremath{\text{TeV}}}   % TeV
\newcommand{\spose}[1]{\hbox to 0pt{#1\hss}}
\newcommand{\ltapprox}{\lesssim}
\newcommand{\gtapprox}{\gtrsim}
\newcommand{\inapprox}{\mathrel{\spose{\lower 3pt\hbox{$\mathchar"218$}}
 \raise 2.0pt\hbox{$\mathchar"232$}}}

\def\SU{{\rm SU}}
\def\SO{{\rm SO}}

\def\U1{{\rm U}(1)}

\newcommand{\4}{\ensuremath{\text{\textbf{4}}}}
\newcommand{\6}{\ensuremath{\text{\textbf{6}}}}

\newcommand{\up}{\uparrow} % spin up
\newcommand{\dn}{\downarrow} % spin down

\begin{document}

%\preprint{}

\title{Baryon spectrum of SU$(4)$ composite Higgs theory with two distinct fermion representations}



\author{Venkitesh Ayyar}
\affiliation{Department of Physics, University of Colorado, Boulder, Colorado 80309, USA}

\author{Thomas DeGrand}
\affiliation{Department of Physics, University of Colorado, Boulder, Colorado 80309, USA}

\author{Daniel C.~Hackett}
\affiliation{Department of Physics, University of Colorado, Boulder, Colorado 80309, USA}

\author{William~I.~Jay}
\affiliation{Department of Physics, University of Colorado, Boulder, Colorado 80309, USA}

\author{Ethan T.~Neil}\email{ethan.neil@colorado.edu}
\affiliation{Department of Physics, University of Colorado, Boulder, Colorado 80309, USA}
\affiliation{RIKEN-BNL Research Center, Brookhaven National Laboratory, \\ Upton, New York 11973, USA}

\author{Yigal~Shamir}
\affiliation{Raymond and Beverly Sackler School of Physics and Astronomy,
Tel~Aviv University, 69978 Tel~Aviv, Israel}

\author{Benjamin Svetitsky}
\affiliation{Raymond and Beverly Sackler School of Physics and Astronomy,
Tel~Aviv University, 69978 Tel~Aviv, Israel}

\date{\today} % change to submission date when posting to arXiv

\begin{abstract}
We use lattice simulations to compute the baryon spectrum of SU(4) lattice gauge theory coupled to dynamical fermions in the fundamental and two-index antisymmetric (sextet) representations simultaneously.
This model is closely related to a composite Higgs model in which the chimera baryon made up of fermions from both representations plays the role of a composite top-quark partner.
The dependence of the baryon masses on each underlying fermion mass is found to be generally consistent with a quark-model description and large-$N_c$ scaling.
We combine our numerical results with experimental bounds on the scale of the new strong sector to derive a lower bound on the mass of the top partner.

\end{abstract}

\maketitle

\begin{flushleft}
%Color coding:

%\vspace{0.5cm}
%\vfill

\end{flushleft}

%\listoffigures
%\listoftables
%\tableofcontents

\section{Introduction}
    \label{sec:intro}
    In this paper we compute the baryon spectrum of SU(4) gauge theory with simultaneous dynamical fermions in two distinct representations, the fundamental \4 and the two-index antisymmetric \6, which is real.
This theory is a slight simplification of a proposed asymptotically free composite Higgs model due to Ferretti~\cite{Ferretti:2013kya,Ferretti:2014qta}---our model contains two Dirac flavors in each representation, while Ferretti's model has five Majorana fermions in the \6 and three Dirac flavors in the \4.
The two essential physical features of Ferretti's model are a composite Higgs boson~\cite{Georgi:1984af,Dugan:1984hq,Contino:2010rs,Bellazzini:2014yua,Panico:2015jxa} and a partially composite top quark~\cite{Kaplan:1991dc}.
The low-energy description of models like Ferretti's has been the subject of recent work; see Refs.~\cite{Golterman:2015zwa,Ferretti:2016upr,Belyaev:2016ftv,Golterman:2017vdj} and references therein.
Composite Higgs scenarios based on other gauge groups are also possible and have been studied recently on the lattice~\cite{Barnard:2013zea,Bennett:2017ttu,Bennett:2017tum,Bennett:2017kbp}.

We have carried out several lattice studies of this interesting model already, including studies of its thermodynamic properties~\cite{Hackett:2017gti,Ayyar:2017uqh,Ayyar:2017vsu} and a detailed calculation of the meson spectrum \cite{Ayyar:2017qdf}; we shall refer to these previous works for a number of technical details.
A preliminary study of the baryon spectrum on a limited set of partially quenched lattices (i.e., with dynamical fundamental fermions but without dynamical sextet fermions) was presented in Ref.~\cite{DeGrand:2016mxr}.

In the context of the Ferretti composite Higgs model, knowing the spectrum of baryon states allows for concrete predictions about their future discovery potential in LHC searches.
One baryon state, made up of valence fermions from both the \textbf{4} and \textbf{6} representations, is of particular interest; in the model it carries the same Standard Model quantum numbers as the top quark, and in fact serves as a top partner, playing a crucial role in the generation of the Higgs potential and of the top-quark mass itself.
We refer to these bound states as ``chimera'' baryons, due to their mixed composition.

Aside from phenomenology beyond the Standard Model, this system offers a rich testing ground for a generalized version of the familiar quark model description of hadronic physics, containing baryons with different expected behavior in the large-$N_c$ limit.
Since baryons in QCD only contain quarks in a single representation, the chimera states are particularly novel from a quark-model perspective.

Our analysis will spend more time on models than is common in modern QCD simulations.
This is an exploratory study of a new system.
There are many baryon states, and it is useful to have an organizing principle to present them.
It is also useful to be able to compare the spectroscopy of this system to that of real-world QCD.
Models are a good way to do that.
The models may also be useful in phenomenology of this and similar theories.

The paper is organized as follows.
In Sec.~\ref{sec:constituent} we introduce a constituent fermion model for the baryons, using large-$N_c$ scaling as an organizing principle.
In Sec.~\ref{sec:lattice} we describe the lattice theory, the ensembles, and the baryonic observables.
In Sec.~\ref{sec:spectrum} we present results for the spectrum and analyze the baryon masses using
global fits to obtain results in the chiral and continuum limits.
Finally, Sec.~\ref{sec:conclusions} summarizes our findings from the perspective of phenomenology
 and presents our conclusions.

Tables containing the various measured quantities have been collected together in
 Appendix~\ref{app:data_tables}.
Technical aspects of the lattice simulation appear in Appendix~\ref{app:lattice}.

%% Clearpage commands are temporary, to make section divisions clearer.
%% Remove in final draft.
%\clearpage
\section{Large-$N_c$ and constituent fermion models}
	\label{sec:constituent}
	\subsection{Baryons in SU$(4)$ with two representations}

Let $N_\4$ and $N_\6$ denote the number of Dirac flavors of fermions in the fundamental and sextet representations.
In the present study $N_\4 = N_\6 = 2$, to be compared with $N_4=3$ and $N_6=5/2$ in Ferretti's model.
The global symmetry group of this model in the massless limit is $ \SU(2 N_\6) \times \SU(N_\4)_L \times \SU(N_\4)_R \times \U1_B \times \U1_A$.
$\U1_B$ is the baryon number of the fundamental fermions, and $\U1_A$ is a conserved axial current.
After spontaneous breaking of chiral symmetry, the unbroken symmetry group is $\SO(2N_\6) \times \SU(N_\4)_V \times \U1_B$.
Additional group theoretical details relating to the fact that the \6 representation is real appear in Ref.~\cite{DeGrand:2015lna}.

The spectrum of the lightest $s$-wave baryons in this theory consists of three classes of states with differing valence fermion content: fundamental-only baryons, sextet-only baryons, and mixed-representation baryons.
Fundamental-only baryons contain four valence fermions and have nonzero $\U1_B$.
We shall denote these bosonic states as $q^4$ states.
Sextet-only baryons contain six valence fermions and we will denote these bosons as $Q^6$ states.
No unique definition of baryon number exists for these pure-sextet objects, although one can single out one of the unbroken $\SO(4)$ generators and call it a baryon number.
In practice we shall only discuss the $Q^6$ states with color indices contracted against the Levi-Civita symbol of $\SO(6)$, as in Ref.~\cite{DeGrand:2015lna}.
[Note that the defining representation of $\SO(6)$ is isomorphic to the \6 of $\SU(4)$.]
Finally, the color-singlet combination of two fundamentals with a single sextet fermion gives a mixed-representation baryon containing three fermions.
We shall denote these fermionic states as $Qqq$ states and refer to them as chimera baryons.

The lightest $Qqq$ chimera baryons are expected to be stable under strong decay, due to conservation of fundamental baryon number $\U1_B$.
The $q^4$ baryons can decay into a pair of chimeras, and a $q^4$ baryon will be stable only if it is sufficiently light.
Because the \textbf{6} of SU(4) is a real representation, di-fermion $QQ$ states live in the same multiplets with fermion-antifermion $\overline{Q}Q$ states;
the same applies to the four-fermion states $QQQQ$, $\overline{Q}QQQ$, etc.
The $Q^6$ states are unstable against decay into three $QQ$ particles, which include in particular some of the states in the Goldstone multiplet of $\SU(2 N_\6) \to \SO(2 N_\6)$ symmetry breaking~\cite{DeGrand:2015lna}.

Mixed-representation baryons represent a new sort of baryon, but the relevant interpolating fields are closely related to those of the QCD hyperons containing a single strange quark (i.e., $\Sigma^*$, $\Sigma$, and $\Lambda$), with the lone sextet fermion playing the role of the strange quark.
From a computational perspective, the only new feature is the presence of an additional color $\SU(4)$ index for the sextet fermion; details appear in Appendix~\ref{app:lattice}.
As in QCD, these mixed-representation baryons are classified according to their total spin $J$ and the isospin $I$ of the $qq$ pair; the three possible states are identified as $\Sigma^\star, (J,I) = (3/2,1)$; $\Sigma, (J,I) = (1/2,1)$; and $\Lambda, (J,I) = (1/2,0)$.
(Total antisymmetry of the operator under exchange of identical fermions forbids a spin-3/2 isosinglet state.)
The chimera analogue of the $\Lambda$ is of particular phenomenological interest, since it plays the role of a partner for the top quark in Ferretti's model.
More information relating to its role as the top partner appears in Sec.~\ref{sec:conclusions} below.

\subsection{Continuum large-$N_c$ expectations}\label{ssec:continuum_large_N}

The properties of both $q^4$ and $Q^6$ baryons have been studied in the continuum
(a partial list of references are
Refs.~\cite{Witten:1979kh,Witten:1983tx,Dashen:1993jt,Dashen:1994qi,Jenkins:1995td,Dai:1995zg,Bolognesi:2006ws}) and on the lattice (in quenched simulations and in ones with a single representation of dynamical fermion---see
Refs.~\cite{DeGrand:2012hd,DeGrand:2013nna,Cordon:2014sda,Appelquist:2014jch,DeGrand:2015lna}).
These states form multiplets in which angular momentum and isospin (flavor) are locked together, $I=J=0,1,\dots N/2$ where $N=4$ or 6 for the $q^4$ and $Q^6$ states.
For the $q^4$ states, this is an aspect of the ``contracted $\SU(2N_c)$" symmetry of large-$N_c$ baryons~\cite{Gervais:1983wq,Gervais:1984rc,Dashen:1993as,Manohar:1998xv}.

Large-$N_c$ predicts that masses of single-representation baryons, which are classified according to their
total spin $J$, should follow a rotor formula.
Mass formulas through $\mathcal{O}(1/N_c)$ for these baryons are given in Refs.~\cite{Adkins:1983ya,Jenkins:1993zu,Buisseret:2011aa}:
\begin{align}\label{eq:srep_mass}
M_B
	&= \dim_r M_r^{(0)} + M_r^{(1)} + B_{rr} \frac{J(J+1)}{\dim_r} \\
	&=
	\begin{cases}
	4 M_4^{(0)} + M_4^{(1)} + B_{44} \frac{J(J+1)}{4} \text{, for } q^4 \\
	6 M_6^{(0)} + M_6^{(1)} + B_{66} \frac{J(J+1)}{6} \text{, for } Q^6,
	\end{cases}
\end{align}
where the dimensions of the representations are $\dim_r = N_c$ for the fundamental and $\dim_r = N_c(N_c-1)/2$ for the two-index antisymmetric representation.
In these expressions, the bulk of the mass of the baryons comes from the leading-order constituent mass term proportional to $\dim_r$.
Sub-leading corrections appear in the term $M_{r}^{(1)}$ and the rotor splitting $B_{rr}$.
Because the $N_c$-dependence has been made explicit, no \emph{a priori} hierarchy is assumed to exist among the parameters $M_{r}^{(0)}$, $M_{r}^{(1)}$ and $B_{rr}$.

Large-$N_c$ together with arguments involving spin-flavor symmetry furnish further predictions for
 mixed-representation baryons~\cite{Dashen:1994qi}.
The key insight is that mixed-representation baryons can be classified according to the (unbroken) flavor symmetry of the fundamental fermions, $\SU(2)_I\times \U1_B$.
With this symmetry, it can be shown that
\begin{align}\label{eq:chimera_mass}
M_{Qqq} = 2 \widetilde{M}_4^{(0)} + \widetilde{M}_6^{(0)} + \widetilde{M}_\text{mix}^{(1)} + \widetilde{B}_{46} \frac{J(J+1)}{\sqrt{24}} + \left(\frac{\widetilde{B}_{44}}{4} - \frac{\widetilde{B}_{46}}{\sqrt{24}} \right) I (I + 1).
\end{align}

Several comments are in order.
First, writing down Eqs.~(\ref{eq:srep_mass}) and~(\ref{eq:chimera_mass}) required no model assumptions beyond large-$N_c$ counting.
The factors of 4 and 6 are conventional.
Second, the tildes in Eq.~(\ref{eq:chimera_mass}) remind us that, from the perspective of large-$N_c$, the expansion parameters of the single-representation baryons are completely unrelated to those of the mixed-representation baryons.
The raw lattice data will soon demonstrate, however, that there is good reason to believe that they are in fact related (\eg, $\widetilde{B}_{44}  \simeq B_{44}$).
Third, each of the parameters above is implicitly a function of the fermion masses $m_4$ and $m_6$.

\subsection{Baryon masses on the lattice}

Motivated by the general arguments above, we now turn to models for describing our lattice data.
Following the methodology we developed studying the meson spectrum of this theory, we express all
 quantities in units of the Wilson flow length scale $\sqrt{t_0}$ \cite{Luscher:2010iy}.
The use of a hat distinguishes these quantities from the corresponding values in lattice units, e.g.,  $\hat{M}_B \equiv (M_B\,a) (\sqrt{t_0}/a)$.
This comes from a mass-dependent scale-setting prescription: the lattice spacing in a given ensemble is derived via direct measurement of $\hat{a} = a/\sqrt{t_0}$ \cite{Ayyar:2017qdf}.

Our simulations are performed across a wide spread of lattice spacings, allowing us to model and remove lattice artifacts.
We expect that the leading-order lattice correction to dimensionless ratios will be linear in the lattice spacing,
\be
\frac{m_1 a}{m_2 a} = \frac{m_1}{m_2} + O(a) +\cdots\,.
\ee
This means that (for each individual state)
\be
\hat{M}_B = \hat{M}_b^0 + A_B \hat a + \cdots\,,
\label{eq:simplest}
\ee
where $\hat{M}_b^0$ is the continuum limit value.

Equation(\ref{eq:simplest}) does not yet include any explicit dependence on the input fermion masses.
One could consider a simple linear dependence on the valence fermion mass $\hat m_v$ (for the $q^4$ or $Q^6$ states),
\be
\hat{M}_B = \hat{M}_b^0 +  \hat{M}_b^1 \hat m_v +  A_B \hat a ,
\label{eq:simplerer}
\ee
or perhaps on both valence and sea masses,
\be
\hat{M}_B = \hat{M}_b^0 +  \hat{M}_b^1 \hat m_v +  \hat{M}_b^2 \hat m_s + A_B \hat a .
\label{eq:simpler}
\ee
In a fit of this form, one would expect $\hat{M}_b^1 > \hat{M}_b^2$, reflecting the fact that the baryon mass depends predominantly on the valence fermion mass.

One could also consider a more complex model based on Eq.~(\ref{eq:srep_mass}), in which all the coefficients have their own lattice artifacts ($\hat B_{rr} = \hat B_{rr}^0 + \hat B_{rr}^1 \hat a$, for example).
In practice, we find that a single artifact term reproduces all the spectroscopy in a multiplet.
Our model for the lattice baryon spectrum is hence
\begin{align}
\hat{M}_{Q^6}
	&= 6 \left[ C_6 + C_{66} \hat{m}_6 \right] + \frac{B_{66}}{6} J (J+1) + A_6 \hat{a}, \label{eq:sextet_mass_model} \\
\hat{M}_{q^4}
	&= 4 \left[ C_4 + C_{44} \hat{m}_4 \right] + \frac{B_{44}}{4} J (J+1) + A_4 \hat{a}, \label{eq:fundamental_mass_model}
\end{align}
\begin{align}
\hat{M}_{Qqq} \label{eq:chimera_mass_model}
	= 2 & \left[ C_4 + C_{44} \hat{m}_4 \right] +  \left[ C_6 + C_{66} \hat{m}_6 \right]  + C_\text{mix} + A_\text{mix} \hat{a}  \nonumber\\
		 &+B_{46} \frac{J(J+1)}{\sqrt{24}} + \left(\frac{B_{44}}{4} - \frac{B_{46}}{\sqrt{24}} \right) I (I + 1).
\end{align}
The constants proportional to $\hat{a}$ are the explicit lattice artifact terms.

It is also worth noting that independent of any fitting, the compatibility of the rotor formula (\ref{eq:srep_mass}) with our baryon mass results can be tested across fermion mass values
with an analog of the Land\'e interval rule: ratios of differences $(M_B(J_1)-M_B(J_2))/(M_B(J_3)-M_B(J_4))$ should be pure numbers, depending only on the $J$'s.
We will present such a test in Sec.~\ref{ssec:no_fits} below as a check on our more elaborate results based on fitting.

The parameters of the lattice models above are related to those of the original large-$N_c$ formulas according to the following relations:
\begin{align}
\widetilde{M}_4^{(0)} &= M_4^{(0)} + M_4^{(1)}/4 = C_4 + C_{44} \hat{m}_4, \\
\widetilde{M}_6^{(0)} &= M_6^{(0)} + M_6^{(1)}/6 = C_6 + C_{66} \hat{m}_6, \\
C_\text{mix} &= \widetilde{M}_\text{mix}^{(1)} - M_4^{(1)}/4 - M_6^{(1)}/6.
\end{align}
This redefinition is desirable from a numerical perspective, since the original large-$N_c$ formulas contain more independent parameters than can be distinguished by data at a single value of $N_c$.
Since the three multiplets of states furnish three linear relations among the constituent masses, a fit can only distinguish between three independent linear combinations of the constituent mass parameters.

%\clearpage
\section{Lattice Theory and Simulation Details}
	\label{sec:lattice}
	
The ensembles used in this study were generated with simultaneous dynamical fermions in the
 fundamental and two-index antisymmetric representations of SU(4).
Each fermion action is a Wilson-clover action, with normalized hypercubic (nHYP) smeared gauge links~\cite{Hasenfratz:2001hp,Hasenfratz:2007rf,DeGrand:2015lna}.
The clover coefficient $c_{SW}$ is set equal to unity for both fermion species.
For the gauge field, we use the nHYP dislocation suppressing (NDS) action, a smeared action designed to reduce gauge-field roughness that would create large fermion forces in the molecular dynamics \cite{DeGrand:2014rwa}.
More details about the action and the gauge configurations can be found in our recent study of
 the meson spectrum~\cite{Ayyar:2017qdf}.

Baryon correlators are noisy, so for the present study, we use only a subset of our full data set, focusing on a dozen ensembles with sufficient statistics to achieve reliable measurements of the baryon spectrum.
Table~\ref{table:ensembles} lists the ensembles used here.
In lieu of repeating the technical details, we summarize some features of these ensembles in Table~\ref{table:ensemble_summary}.

All the ensembles in the present study have volume $V=16^3 \times 32$.
In our previous study, we estimated the finite volume effects for mesons and concluded that they were at the level of a few percent.
Baryons, of course, are a different story, because their sizes are expected to be larger than those of mesons.
Still, the pseudoscalar decay constants in the SU(4) gauge theory are larger than those of SU(3) and, since finite volume corrections from pion loops are proportional to $1/F_{P}^2$, we expect that they are smaller than in SU(3)
(see Ref.~\cite{Ayyar:2017qdf} for a discussion).
Sec.~\ref{ssec:fit_results} presents a preliminary estimate of finite-volume effects in our analysis.

\begin{table}[t]
\centering
\setlength{\tabcolsep}{12pt} % "Wider" columns
\begin{tabular}{ l l l }
				& min 	& max \\
\toprule
$t_0/a^2$			& 1.06 	& 2.67 \\
$M_{P4}/M_{V4}$	& 0.55	& 0.79 \\
$M_{P6}/M_{V6}$	& 0.47	& 0.73 \\
$M_{p4}L$		& 4.23	& 8.16 \\
$M_{p6}L$		& 4.03	& 8.91 \\
\end{tabular}\caption{Summary of basic physical properties of the ensembles used in this study.}
\label{table:ensemble_summary}
\end{table}

We extract baryon masses $M_B$ in the usual way from two-point correlation functions.
We shall often consider the baryon masses as functions of the fermion masses $m_4$ and $m_6$, 
which are defined by the axial Ward identity (AWI), 
\begin{align}
\partial_\mu \left\langle 0 \right| A_{\mu a}^{(r)}(x) \mathcal{O}_{r} \left| 0 \right \rangle = 
2 m_r \left\langle 0 \right| P_a^{(r)} (x) \mathcal{O}_r \left| 0 \right \rangle,
\end{align}
where $a$ is an isospin index.
Table~\ref{table:fermion_masses} collects the measured fermion masses,
 taken from Ref.~\cite{Ayyar:2017qdf}.
Further information about our conventions and methods for spectroscopy 
appears in Appendix~\ref{app:lattice}.

%\clearpage
\section{Spectrum Results and Analysis}
	\label{sec:spectrum}
	\subsection{Results from the raw data}\label{ssec:no_fits}

\begin{figure}[t]
\includegraphics[width=0.8\textwidth]{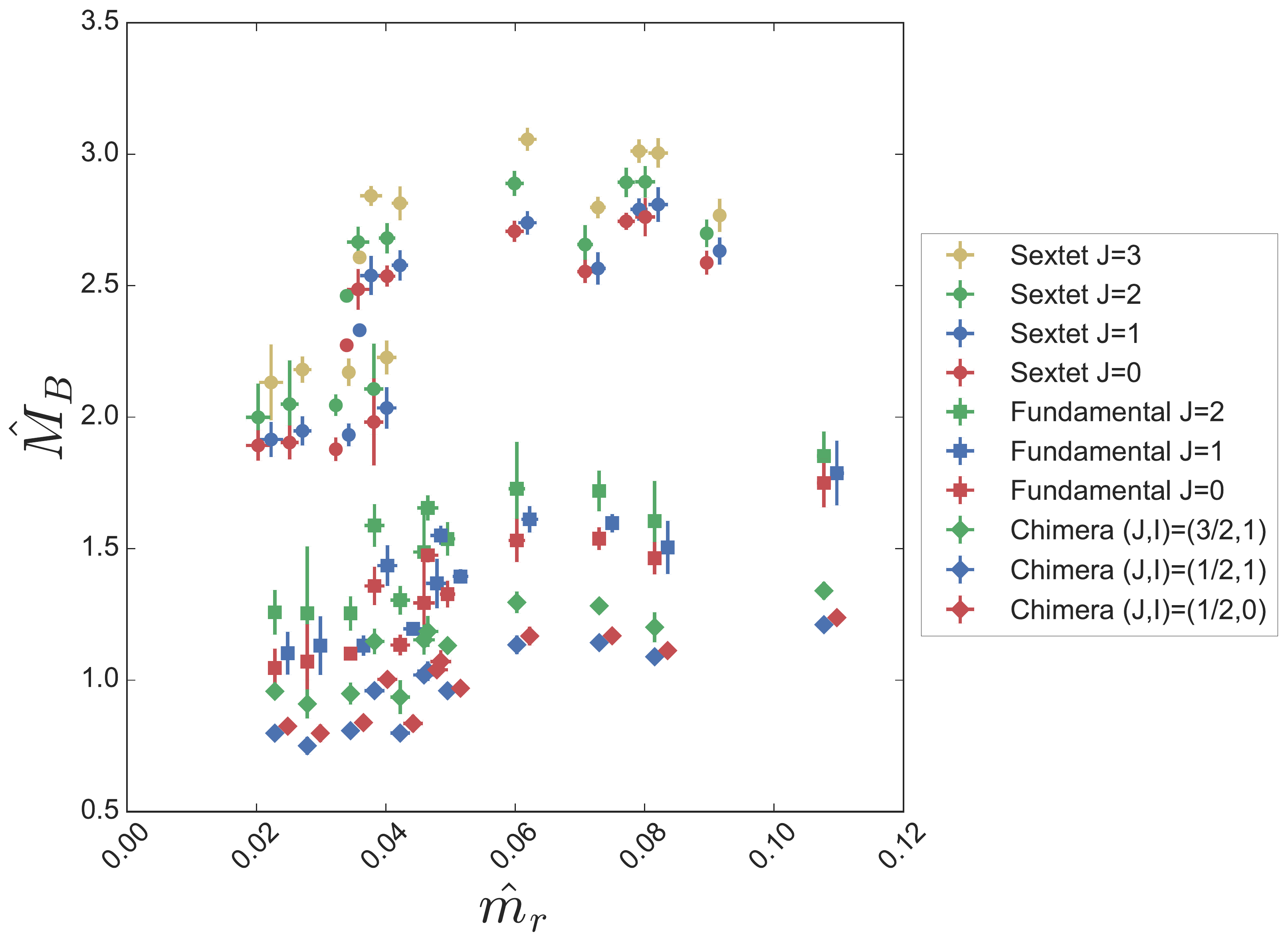}
\caption{
	Lattice data for the baryon mass spectrum $\hat{M}_B$.
	The horizontal positions contain small offsets to reduce overlap and aid readability.
	\label{fig:raw_baryon_data}
}
\end{figure}

The tables containing the measured values for the baryons have been collected together in Appendix \ref{app:data_tables}.
The figures in this section summarize the content of the tables.
Figure~\ref{fig:raw_baryon_data} shows the measured spectrum of baryon masses.
Masses of single-representation baryons $\hat{M}_{r}$ are plotted as functions of the corresponding AWI fermion mass $\hat{m}_r$.
The chimera baryons are plotted as a function of $\hat{m}_4$, although one expects some dependence on $\hat{m}_6$ as well.
The baryon masses all increase with fermion mass, but no clear functional dependence is conspicuous.
As in our meson study, lattice artifacts---which we shall model and
remove---obscure the underlying linear nature of our data.
To motivate the forthcoming analysis in Sec.~\ref{ssec:fit_results}, we first consider
 evidence for the models which exists \emph{before} fitting.

\begin{figure}[htb]
\includegraphics[width=1.0\textwidth]{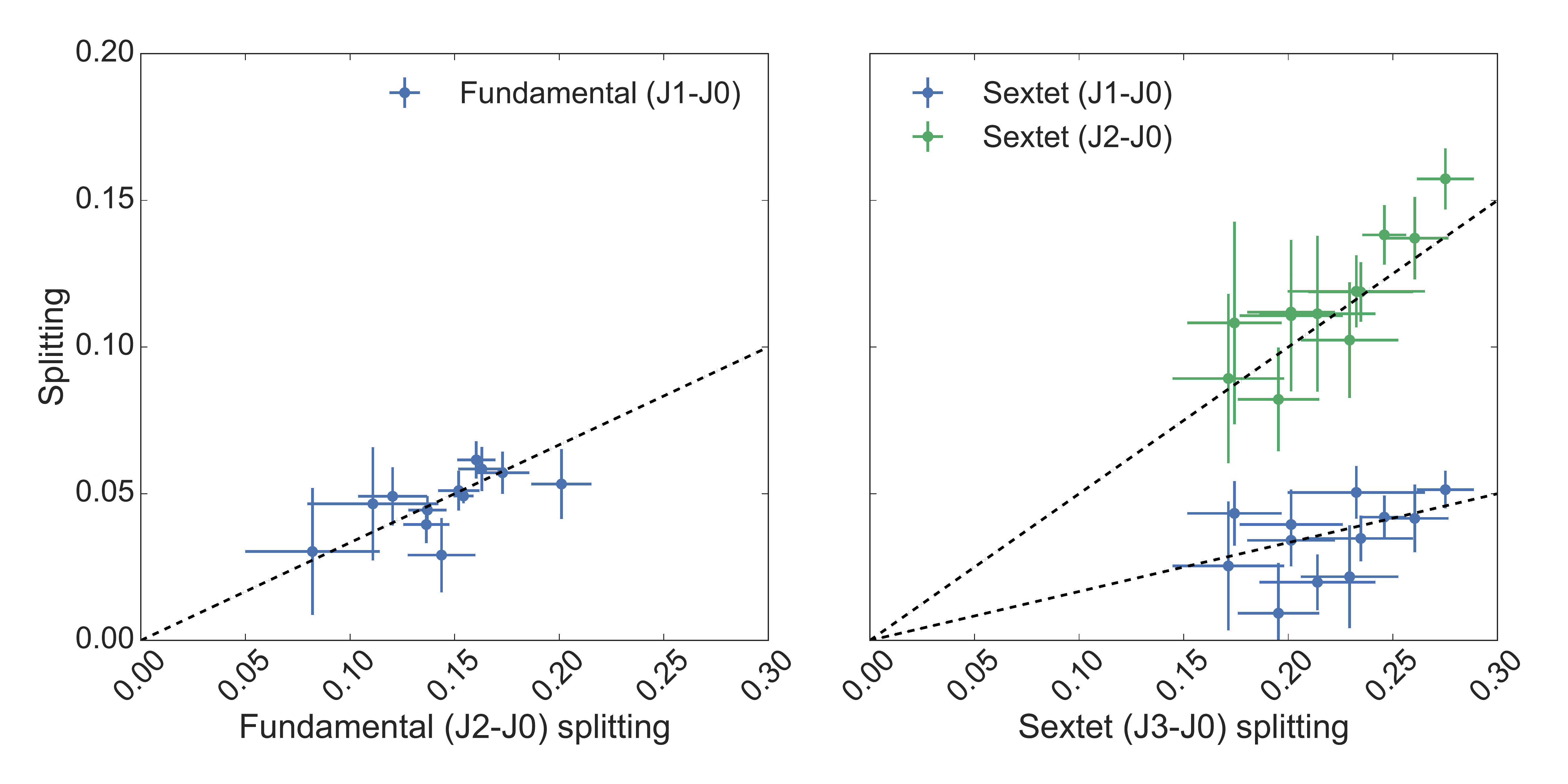}
\caption{
	Mass splittings between baryons in the fundamental (left) and sextet (right) representations.
	The lines indicate the expected $J(J+1)$ behavior.
	\label{fig:srep_splittings}
}
\end{figure}

According to the large-$N_c$ model of Eq.~(\ref{eq:srep_mass}), ratios of baryon mass differences for the $q^4$ or $Q^6$ states are parameter-free functions of the spins.
In particular, the parameter $B_{rr}$ only controls the overall size of the splittings.
Figure~\ref{fig:srep_splittings} shows mass differences among the single-representation baryons, with errors on the differences from a jackknife.
The dotted lines are the expected values from $J(J+1)$ splittings.
The rotor behavior is clearly evident in the raw data.

\begin{figure}[htb]
\includegraphics[width=0.5\textwidth]{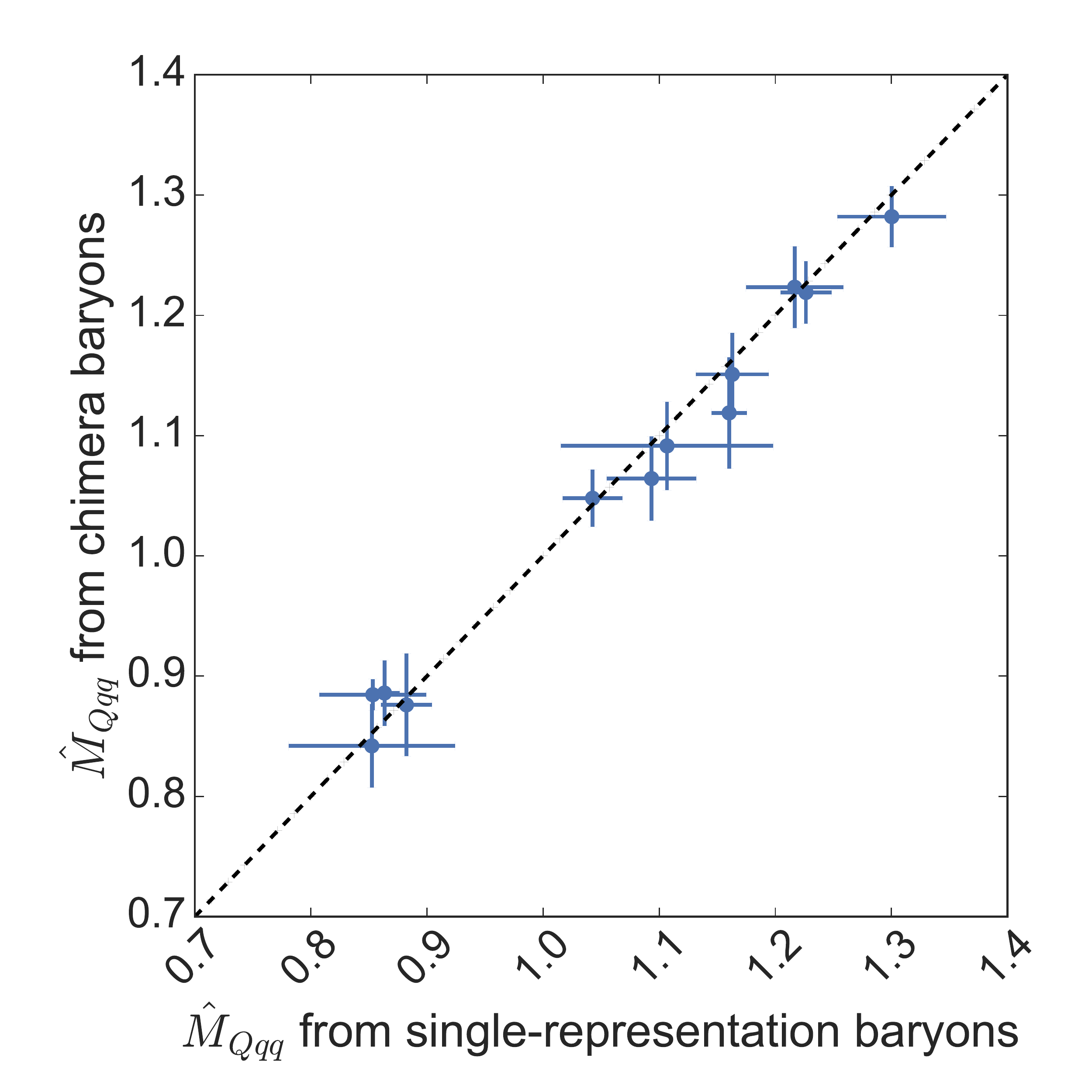}
\caption{
	Estimates for the total constituent mass of the chimera baryons from the chimeras themselves (vertical axis) again the estimates from single-representation baryons (horizontal axis).
	The line indicates the expectation that the two independent estimates agree.
	\label{fig:chimera_constituent}
}
\end{figure}

For single-representation baryons, the mass of the $J=0$ state furnishes estimates for the individual constituent masses.
The constituent mass of a chimera baryon is therefore nearly
\begin{align}
\hat{M}_{Qqq,\text{constituent}} \simeq \frac{\hat{M}_{q^4}^{(J=0)}}{2} + \frac{ \hat{M}_{Q^6}^{(J=0)}}{6}.
\end{align}
On the other hand, one can use Eq.~(\ref{eq:chimera_mass}) to eliminate the splitting terms in favor of the spin independent
contribution by averaging the chimera baryon masses together with appropriate weights,
\begin{align}
\hat{M}_{Qqq,\text{constituent}} \simeq\left( 2 \hat{M}_{Qqq}^{(J,I)=(3/2,1)} + \hat{M}_{Qqq}^{(J,I)=(1/2,1)} + \hat{M}_{Qqq}^{(J,I)=(1/2,0)} \right) / 4.
\end{align}
Figure~\ref{fig:chimera_constituent} shows these two raw-data estimates for the total constituent mass of the chimera baryons plotted against each other, with errors from a jackknife.
The line indicates equality, as predicted by the large-$N_c$ model.
The impressive agreement of the two estimates suggests that the chimera baryons should be modeled together with the single-representation baryons.

\begin{figure}[htb]
\includegraphics[width=1.0\textwidth]{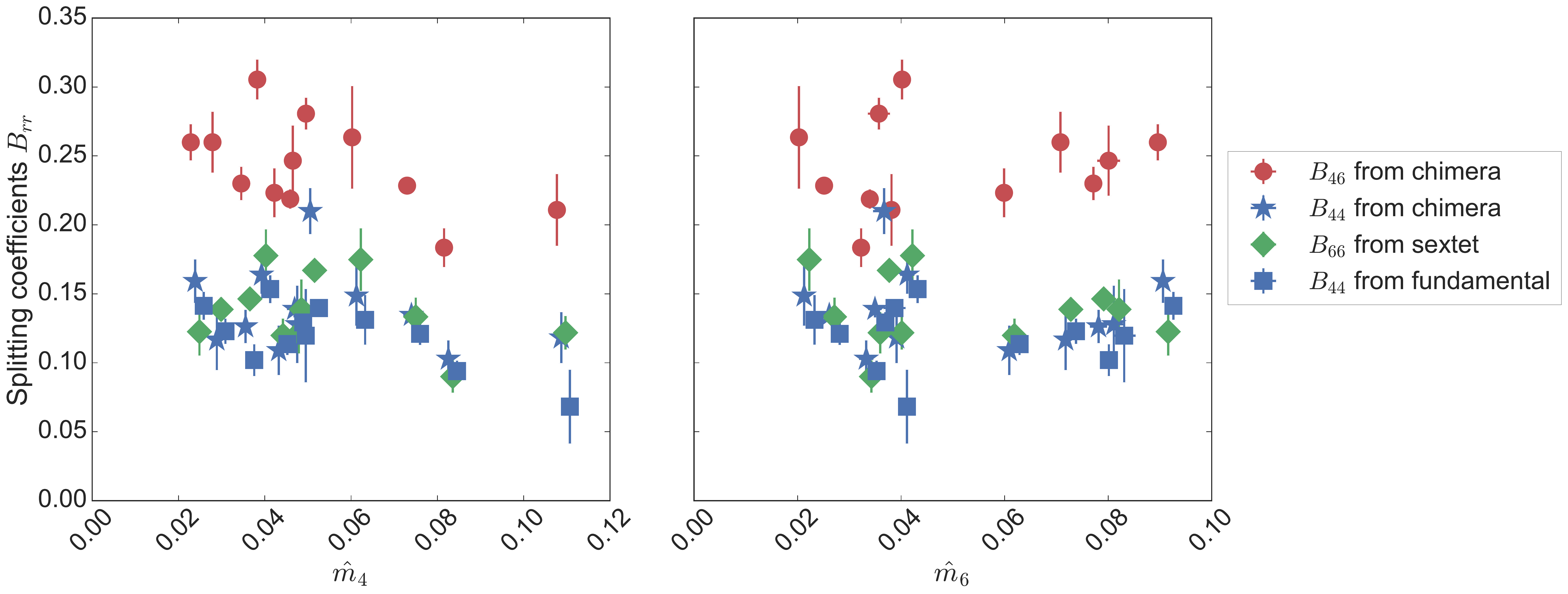}
\caption{
	Estimates of the splitting coefficients $B_{44}$, $B_{66}$, and $B_{46}$ on each ensemble as functions of $\hat{m}_4$ (left) and $\hat{m}_6$ (right).
	The horizontal positions contain small offsets to reduce overlap and aid readability.
	\label{fig:splitting_coefficients}
}
\end{figure}

Finally, we examine the strength of the splitting terms.
Equation (\ref{eq:chimera_mass_model})  says that
\begin{align}
B_{44} &= \frac{2}{3} \left( 2 \hat{M}_{Qqq}^{(J,I)=(3/2,1)} + \hat{M}_{Qqq}^{(J,I)=(1/2,1)} - 3 \hat{M}_{Qqq}^{(J,I)=(1/2,0)} \right), \\
B_{46} &= \frac{\sqrt{24}}{3} \left( \hat{M}_{Qqq}^{(J,I)=(3/2,1)} - \hat{M}_{Qqq}^{(J,I)=(1/2,1)} \right).
\end{align}
Likewise, the single representation formulas (\ref{eq:sextet_mass_model}) and (\ref{eq:fundamental_mass_model}) say that
\begin{align}
B_{44} &= \frac{2}{3} \left(\hat{M}_{q^4}^{(J=2)} - \hat{M}_{Q^6}^{(J=0)} \right), \\
B_{66} &= \frac{1}{2} \left(\hat{M}_{Q^6}^{(J=3)} - \hat{M}_{Q^6}^{(J=0)} \right).
\end{align}
Figure~\ref{fig:splitting_coefficients} shows these estimates for the rotor splitting coefficients $B$, displayed as functions of $\hat{m}_4$ ($\hat{m}_6$) in the left (right) pane, with errors from a jackknife.

Physically-motivated models of baryons predict different mass dependence for the splitting coefficients.
For instance, if the $J(J+1)$ term arises from rigid rotation (e.g., of a skyrmion),
the coefficient should scale inversely with the mass of the baryon \cite{Adkins:1983ya}.
In non-relativistic quark models, the splittings arise from a color hyperfine interaction
and (in analogy with the familiar hyperfine interaction of atomic physics) scale
inversely with the square of the constituent quark mass~\cite{DeRujula:1975qlm}.
The raw data show considerable spread, and no particular functional dependence is
evident for any of the $B$ coefficients in either pane.
The models (\ref{eq:sextet_mass_model}), (\ref{eq:fundamental_mass_model}),
and~(\ref{eq:chimera_mass_model}) therefore treat the $B$ coefficients as constants.
It is worth noting that the two independent estimates of $B_{44}$ from the
fundamental baryons (squares) and from the chimera baryons (stars) are consistent
within the uncertainty of the data.

Another feature of Fig.~\ref{fig:splitting_coefficients} is the large value of $B_{46}$ compared to the other splitting coefficients.
This is easily understood in a model where the spin splittings are due to one-gluon exchange, that is, $V_{ij} \propto C_{ij}\mathbf{S}_i\cdot \mathbf{S}_j$ where $C_{ij}$ is a color factor and $\mathbf{S}_i$ is the spin of constituent $i$.
The appropriate color factors are $C_{qq} = 5/8$ for the $q^4$ baryons (and for the $qq$ diquark in the $Qqq$ baryon), $C_{QQ} = 1/2$ for the $Q^6$ baryons, and $C_{Qq} = 5/4$ for the mixed interaction chimeras.
In other words, one expects $B_{46} / B_{44} \sim C_{Qq} / C_{qq}$ and $B_{46} / B_{66} \sim C_{Qq} / C_{QQ}$.
This expectation is in agreement with the qualitative behavior of Fig.~\ref{fig:splitting_coefficients}, which suggests that $B_{46}$ is roughly twice as large as $B_{44}$ or $B_{66}$.

\subsection{Fitting mass and lattice-spacing dependence}\label{ssec:fit_results}

\begin{figure}[htb]
\includegraphics[width=0.8\textwidth]{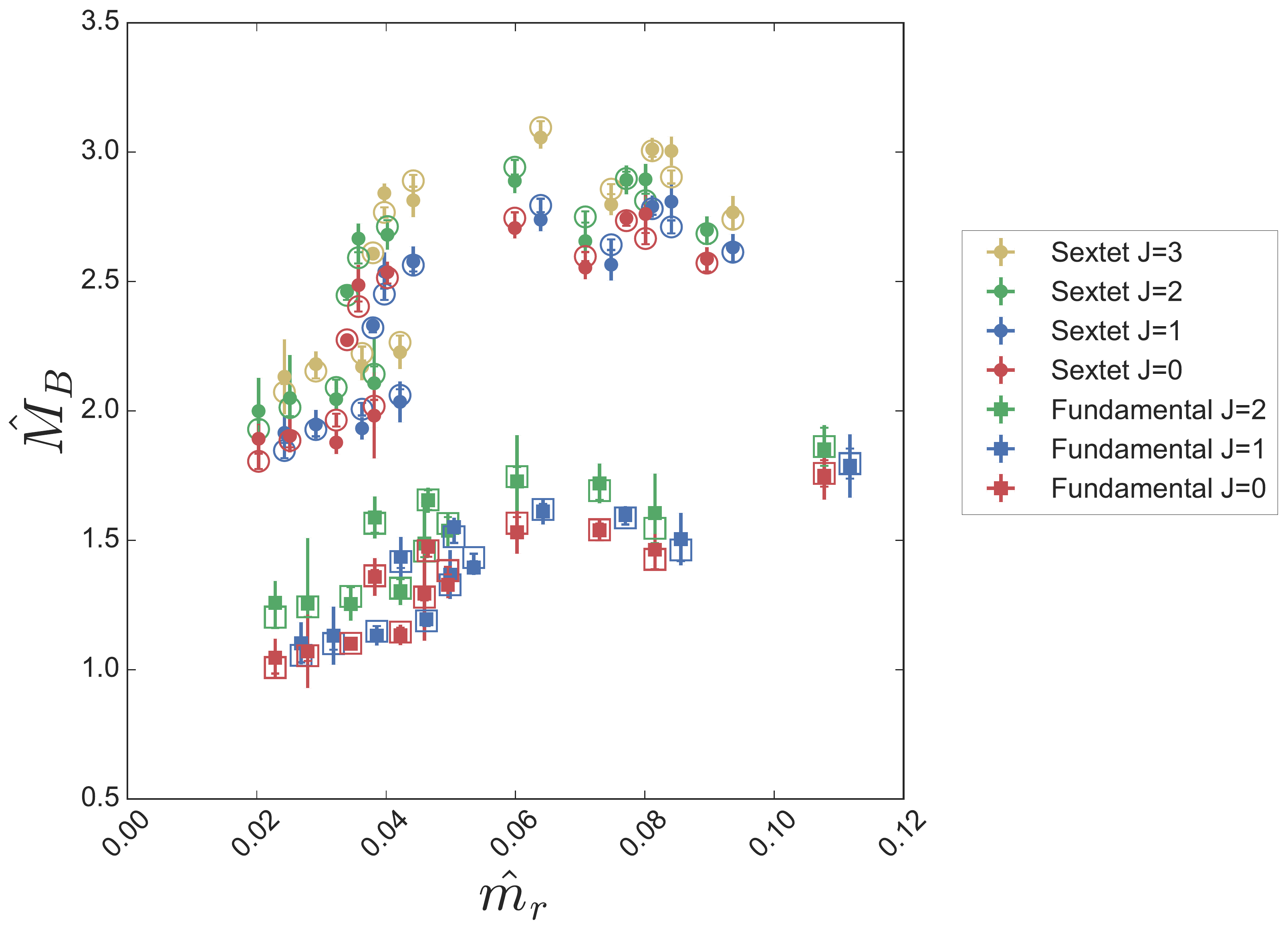}
\caption{
	Data (solid marker) and linear fits (open marker) to each state in the fundamental and sextet multiplets using the simple linear model (\ref{eq:simplest}), including a linear term proportional to $\hat{a}$.
	The error bars of the open markers are those of the fit.
	The horizontal positions contain small offsets to reduce overlap and aid readability.
	\label{fig:basic_linear_fit}
}
\end{figure}

We now present fit results modeling the dependence of our baryons on the fermion masses and lattice spacing, as outlined in Sec.~\ref{sec:constituent}.
To justify the assumption that lattice artifacts are proportional to $\hat{a}$ in our general model, we begin by conducting a simple linear fit following Eq.~(\ref{eq:simplest}) for each state individually in the fundamental and sextet multiplets.
Figure~\ref{fig:basic_linear_fit} shows the result of these fits, which are in excellent agreement with the data.
The parameter corresponding to the artifact is approximately the same for all the states within a given multiplet, giving support to the models of Eqs.~(\ref{eq:sextet_mass_model})--(\ref{eq:chimera_mass_model}).
The case for the chimera baryons is similar, using the form (\ref{eq:simplerer}) which is linear in both $\hat{m}_4$ and $\hat{m}_6$.

\begin{figure}[t]
	\includegraphics[width=0.75\textwidth]{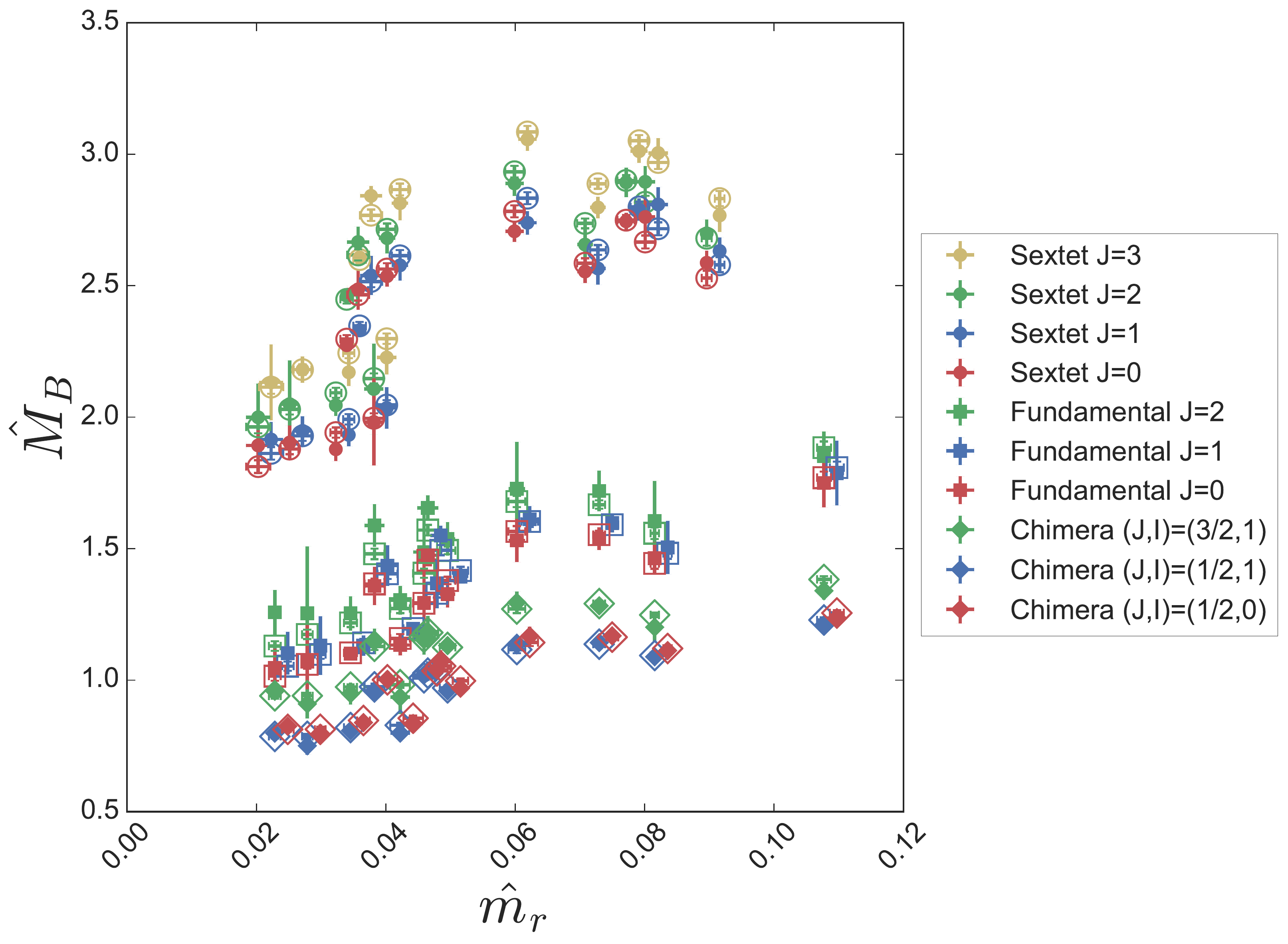}
	\caption{
		Results from the joint correlated fit of all baryon data to Eqs.~(\ref{eq:sextet_mass_model}), (\ref{eq:fundamental_mass_model}), and~(\ref{eq:chimera_mass_model}).
		The data (solid marker) correspond well to the fit (open marker) at each point.
		The errors bars of the open markers are those of the fit.
		The horizontal positions contain small offsets to reduce overlap and aid readability.
		\label{fig:raw_baryon_data_fit_overlay}
	}
\end{figure}

Based on the success of these simple fits, we proceed to a simultaneous global fit.
Using Eqs.~(\ref{eq:sextet_mass_model})--(\ref{eq:chimera_mass_model}), we simultaneously model all 10 baryon masses on 12 ensembles.
Single-elimination jackknife furnishes errors and correlations among the masses.
The model used contains 11 free parameters, leaving $120-11=109$ degrees of freedom.
The resulting fit is good, with $\chi^2/\text{DOF} \simeq 93/ 109 = 0.85$.
Figure~\ref{fig:raw_baryon_data_fit_overlay} shows the data with the fit overlaid.

\begin{figure}[t]
	\includegraphics[width=1.0\textwidth]{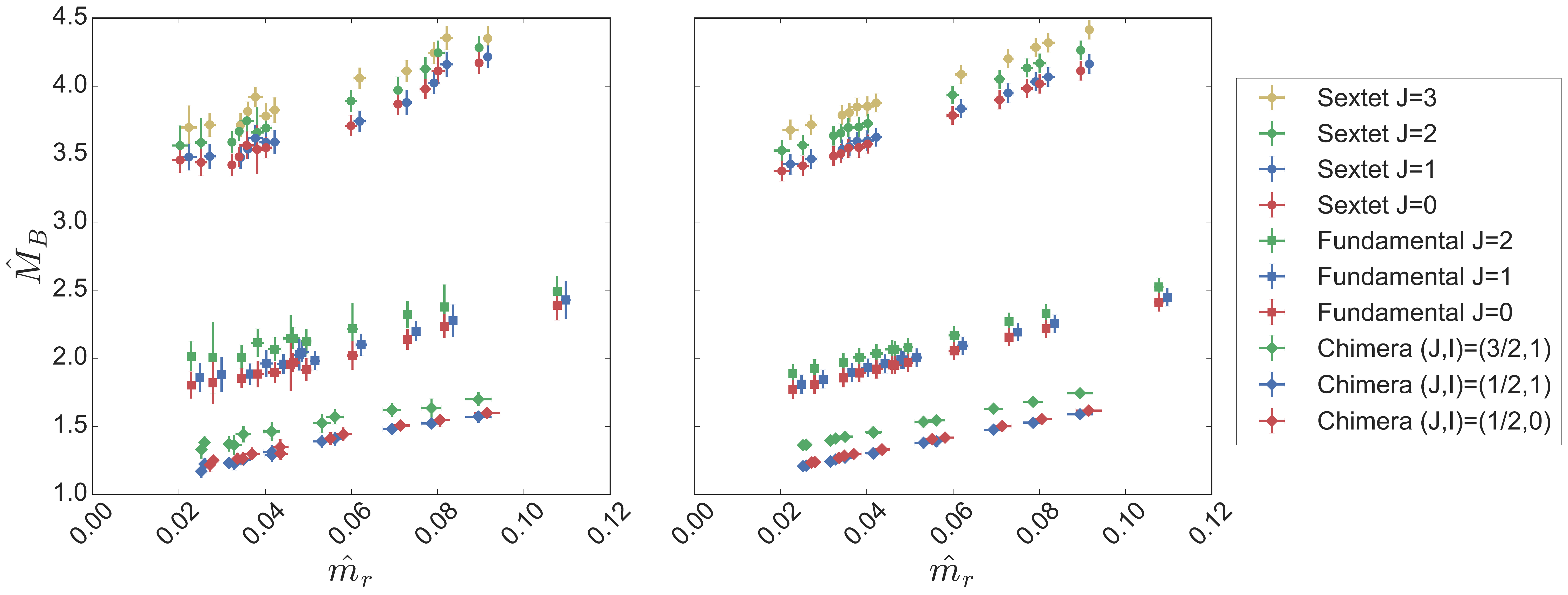}
	\caption{
		Results from the joint correlated fit of all baryon data to  Eqs.~(\ref{eq:sextet_mass_model})--(\ref{eq:chimera_mass_model}).
		The left pane shows continuum masses obtained by subtracting the lattice artifact from the data.
		The right pane shows the corresponding fit, with the lattice artifact term removed.
		The horizontal positions contain small offsets to reduce overlap and aid readability.
		\label{fig:fit_result_ctm}
	}
\end{figure}

Figure~\ref{fig:fit_result_ctm} shows the same fit result after subtracting the lattice
 artifacts (proportional to $\hat{a}$) from each state.
In this figure, the sextet baryons are plotted as functions of $\hat{m}_6$, while the
 fundamental baryons are plotted as functions of $\hat{m}_4$.
Because the masses of the chimera baryons are joint functions of the fermion mass in both representations,
they are plotted against the combination $(2 C_{44} \hat{m}_4 + C_{66} \hat{m}_6)/6$.
The fit formula is linear in this combination.
(The factor of 6 is arbitrary and chosen to give the independent variable a similar range to $\hat{m}_4$ and $\hat{m}_6$.)
The underlying linear behavior for all the baryons is now clearly visible.

In general, one expects the masses of the single-representation baryons to depend predominantly on the mass of the valence fermions in the same representation and only weakly on the sea fermions in the other representation.
The analogy in QCD is the mass of the proton, which also receives virtual contributions from strange quarks.
The fact that our models produce a good fit while neglecting these effects suggests that they are small, although we expect their existence as a matter of principle.
Repeating the fits including sea dependence did not produce any important changes in the results.

As a preliminary estimate of finite-volume effects, we repeated the above fitting analysis keeping only ensembles with $M_P L > (4.25, 4.5, 4.75, 5.0)$.
All qualitative features of the spectrum---the ordering, the rotor splitting, and general placement of the states---were stable against these variations.
Quantitatively, the fit parameters were unchanged at the level of roughly one standard deviation.
Because the focus of the present study is largely qualitative in nature, we leave a more systematic study of these effects for future work.

\subsection{Physical limits}

\begin{figure}[t]
\includegraphics[width=0.75\textwidth]{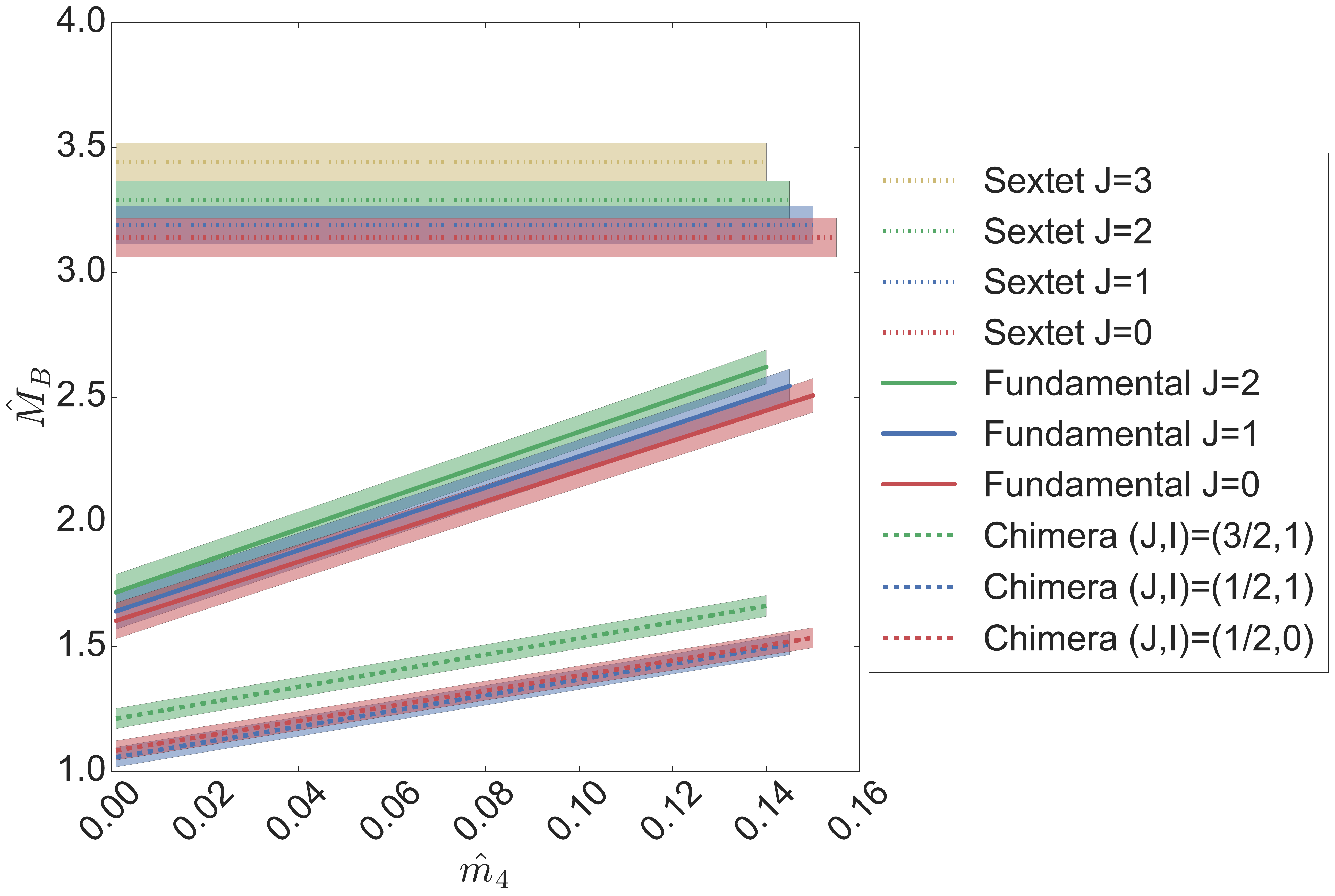}
\caption{
	The baryon spectrum in the $\hat{m}_6 \rightarrow 0$ limit.
	\label{fig:ferretti_limit}
}
\end{figure}

The fit results of the previous section are most interesting in two limits:
the $\hat{m}_6 \rightarrow 0$ chiral limit and the double limit $\hat{m}_4, \hat{m}_6 \rightarrow 0$.
The former limit is important in Ferretti's model, where the Higgs boson arises
 (before perturbative coupling to the Standard Model) as an exactly massless sextet Goldstone boson.
Figure~\ref{fig:ferretti_limit} shows the baryon spectrum in the $\hat{m}_6 \rightarrow 0$ limit,
displayed as a function of the fundamental fermion mass.
By construction, the masses of the sextet baryons are independent of $\hat{m}_4$.
Likewise, the masses of  fundamental and chimera baryons are linear in the fundamental fermion mass.
The lightest baryons in the spectrum are the nearly-degenerate $J=1/2$ chimera baryons,
 the analogues of the $\Sigma$ and $\Lambda$ in QCD.
Regarding these two states, it is interesting to note that we observe an inverted
 multiplet $M_\Lambda \gtrsim M_\Sigma$ with respect to the ordering in QCD, where $M_\Sigma > M_\Lambda$.
This ordering is present in the raw lattice data on all the ensembles we considered.
Using a non-relativistic quark model as a guide, one would also expect this inversion to occur in QCD if the strange
quark were lighter than the up and down quarks.

The spectrum in the double chiral limit ($\hat{m}_4, \hat{m}_6 \rightarrow 0$) corresponds to
the vertical axis in Fig.~\ref{fig:ferretti_limit}.
For convenience, Table~\ref{table:dual_chiral_limit} also records numerical values for the
spectrum in the double chiral limit, both in units of the flow scale $\sqrt{t_0}$ and in units
of the sextet pseudoscalar decay constant, which we determined in our previous study of the meson spectrum.

\begin{table}[t]
\centering
\setlength{\tabcolsep}{12pt} % "Wider" columns
	\begin{tabular}{lll}
						& $M_B \sqrt{t_0}$ &  $M_B / F_6$ \\
	\toprule
	Fundamental vector	meson &	0.74(3) & 4.2(3)\\
	Sextet vector meson	 	&	0.80(3) & 4.6(3)\\
	\hline
	Chimera $(J,I)=(1/2,0)$ 	&      1.08(4) &   6.4(4) \\
	Chimera $(J,I)=(1/2,1)$ 	&      1.05(4) &   6.2(4) \\
	Chimera $(J,I)=(3/2,1)$ 	&      1.21(4) &   7.1(5) \\
	Fundamental $(J=0)$ 	&      1.60(7) &   9.4(7) \\
	Fundamental $(J=1)$ 	&      1.63(7) &   9.6(7) \\
	Fundamental $(J=2)$ 	&      1.71(7) &  10.1(7) \\
	Sextet $(J=0)$ 			&      3.14(8) &  18(1) \\
	Sextet $(J=1)$ 			&      3.19(8) &  19(1) \\
	Sextet $(J=2)$ 			&      3.29(8) &  19(1) \\
	Sextet $(J=3)$ 			&      3.44(8) &  20(1) \\
	\end{tabular}
	\caption{
	The baryon spectrum in the double chiral limit ($\hat{m}_4, \hat{m}_6 \rightarrow 0$) in units of the flow scale $\sqrt{t_0}$ and of the sextet pseudoscalar decay constant $F_6$.
	For comparison, the masses of the fundamental and sextet vector mesons in this limit are also included.
	Mesonic quantities were determined in~\cite{Ayyar:2017qdf}.
	}
	\label{table:dual_chiral_limit}
\end{table}

\subsection{Scalar matrix element}

We can repurpose our calculations of the mass dependence of the baryons to extract the scalar matrix element $\left\langle B \right| \bar{\psi} \psi \left| B \right\rangle$ using the Feynman-Hellmann theorem \cite{Junnarkar:2013ac}.  In the context of composite dark matter models, this matrix element determines the coupling of the Higgs boson to the dark matter and is thus required to calculate the cross section for dark matter direct detection.
(For a review of composite dark matter models, see Ref.~\cite{Kribs:2016cew}.)
Following Ref.~\cite{Appelquist:2014jch}, we define the quantity $f_r^B$ for the lowest-lying baryon in each representation $r$:
\begin{align} \label{eq:scalar_matrix_elt}
f_r^{B} \equiv \frac{\hat{m}_r}{\hat{M}_r} \frac{\partial \hat{M}_r}{\partial \hat{m}_r} = \frac{m_r}{M_r} \left\langle B \right| \bar{\psi} \psi \left| B \right\rangle,
\end{align}
The dimensionless factor $\hat{m}_r/\hat{M}_r$ serves to cancel the dependence on the renormalization prescription of $\bar\psi \psi$ in this expression.
We expect $f_r^B$ to be equal to zero in the chiral limit, and to approach unity in the heavy fermion ($\hat{m}_r \rightarrow \infty)$ limit.

\begin{figure}[t]
\includegraphics[width=1.0\textwidth]{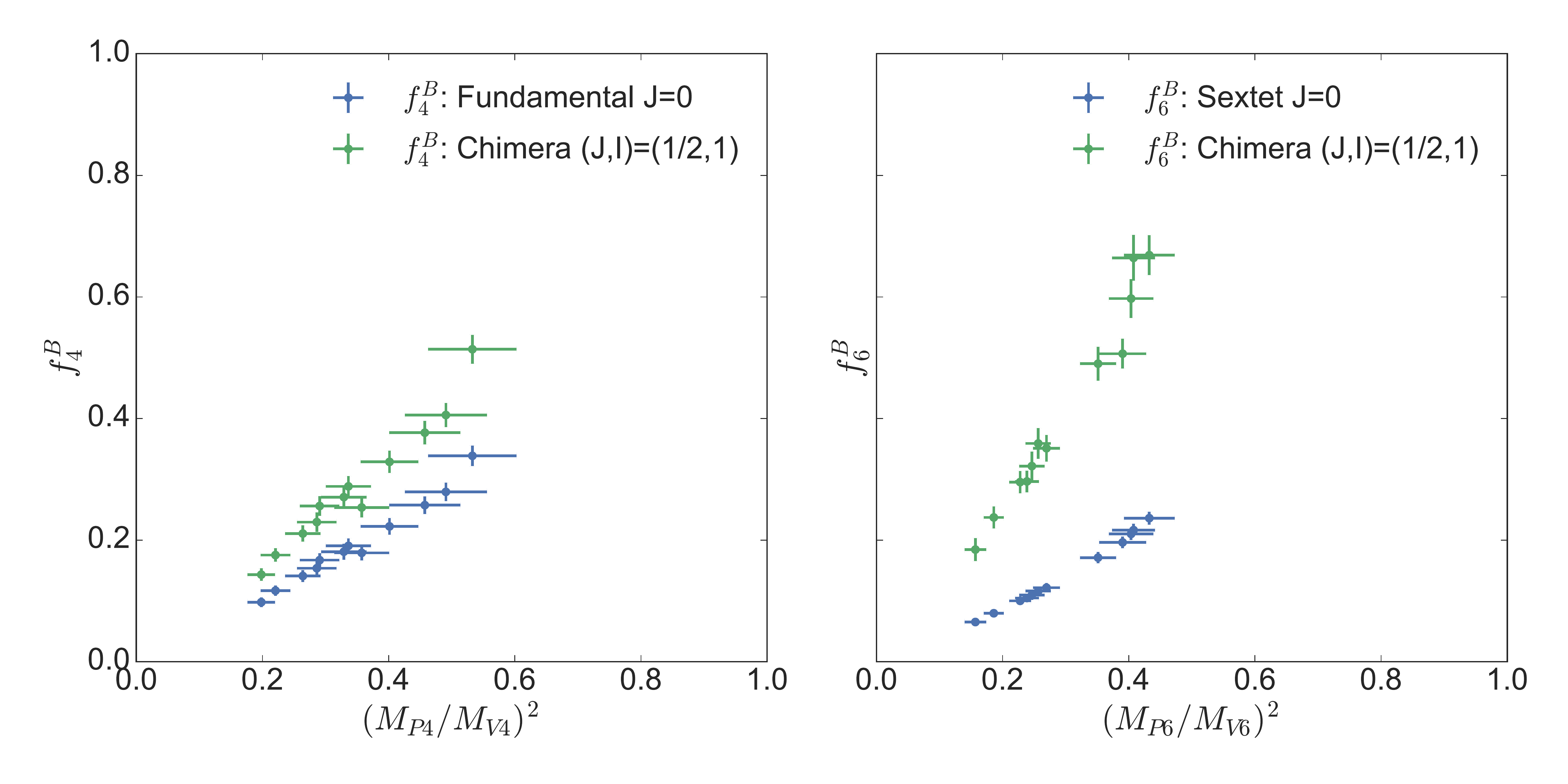}
\caption{
	Baryon matrix elements of the scalar density, defined via Eq.~(\ref{eq:scalar_matrix_elt}).
	Only the lightest state in each representation is shown; the heavier states are similar.
	The mesonic quantities were determined in~\cite{Ayyar:2017qdf}.
	\label{fig:scalar_matrix_elt}
}
\end{figure}

Figure~\ref{fig:scalar_matrix_elt} shows our results for $f_r^B$ in the fundamental and sextet representations, displayed as functions of the pseudoscalar-to-vector mass ratios on each ensemble.
Only the values for the lightest state in each representation are shown; the heavier states are similar.
We note that, in the Ferretti model, none of these baryon states plays the role of a dark matter candidate because of the precise Standard Model charge assignments of the model.
However, it is interesting that the values for $f_r^B$ for the fundamental and sextet baryons resemble results seen previously in a number of different gauge theories (in particular, see Fig.~4 of Ref. \cite{DeGrand:2015zxa}).
The values of both rescaled matrix elements for the chimera baryon come out significantly larger.

\clearpage
\section{Discussion}
    \label{sec:conclusions}
    In this paper we have described the baryon spectrum of SU(4) gauge theory coupled to dynamical fermions in the \textbf{4} and \textbf{6} representations.
Our simulations of this theory with fully dynamical fermions in multiple representations are the first of their kind.
Our choice of this theory was inspired by its close similarity to an asymptotically free composite Higgs model first studied by Ferretti \cite{Ferretti:2014qta}.

The baryon spectrum of this theory contains three classes of baryons with differing valence fermion content:
fundamental-only baryons, sextet-only baryons, and mixed-representation baryons.
Our analysis began by considering raw lattice results for the baryon masses to motivate a joint model based on large-$N_c$ counting.
The important features of this model---a $J(J+1)$ rotor behavior for splittings and shared set
 of constituent masses for fermions---were clearly visible even before fitting.
The resulting fit was successful and identified a significant lattice artifact proportional to the lattice spacing in each baryon multiplet.

After removing the lattice artifact, the baryon mass data show linear dependence on the fermion mass.
Presumably, a more careful analysis in the spirit of heavy baryon chiral perturbation theory \cite{Jenkins:1990jv,Bernard:2007zu} would predict non-analytic behavior similar to that of QCD.
The precision of our data does not allow us to test these predictions.
Observing this behavior in QCD is notoriously difficult, requiring very light fermions.

In Ferretti's model, the Standard Model top quark mixes linearly with the analogue of the $\Lambda$.
This happens because the fundamental fermions carry $\SU(3) \times \SU(3)' \supset \SU(3)_c$ flavor quantum numbers and transform as a $\bar{\textbf{3}}$, while sextet fermions are uncharged under $\SU(3)_c$.
The fundamental fermions within the top partner are contracted anti-symmetrically to form a \textbf{3} of $\SU(3)_c$.
Discarding one of the three fundamental fermions as we did in this paper, we obtain a $qq$ state, still anti-symmetrized on its flavor index and hence an isospin singlet.
Because the $qq$ state is antisymmetric on both its $\SU(4)$-color and flavor indices, the spins must couple anti-symmetrically as well into a $J_{qq}=0$ state.
Thus, the top partner is the analogue of the $\Lambda$ hyperon in QCD.

The phenomenology of composite Higgs completions of the Standard Model is commonly presented in terms of a ratio $\xi = v^2/F^2$ where $v$ is the Higgs vacuum expectation value (246 GeV), and $F$ is the relevant pseudoscalar decay constant.
In the Ferretti model, $F=F_6/\sqrt{2}$ where $F_6$ is the decay constant of the sextet Goldstone bosons. (The factor of $\sqrt{2}$ is due to our normalization convention, which corresponds to $F_\pi \simeq 130$ MeV in QCD.)
In the absence of a direct detection of new resonances, a discovery of new physics can come through a deviation of some observable from its Standard Model value, which would point to a value of $\xi$ and hence of $F_6$.
That would set the scale for other hadronic observables in the new physics sector.

\begin{figure}[t]
\includegraphics[width=1.0\textwidth]{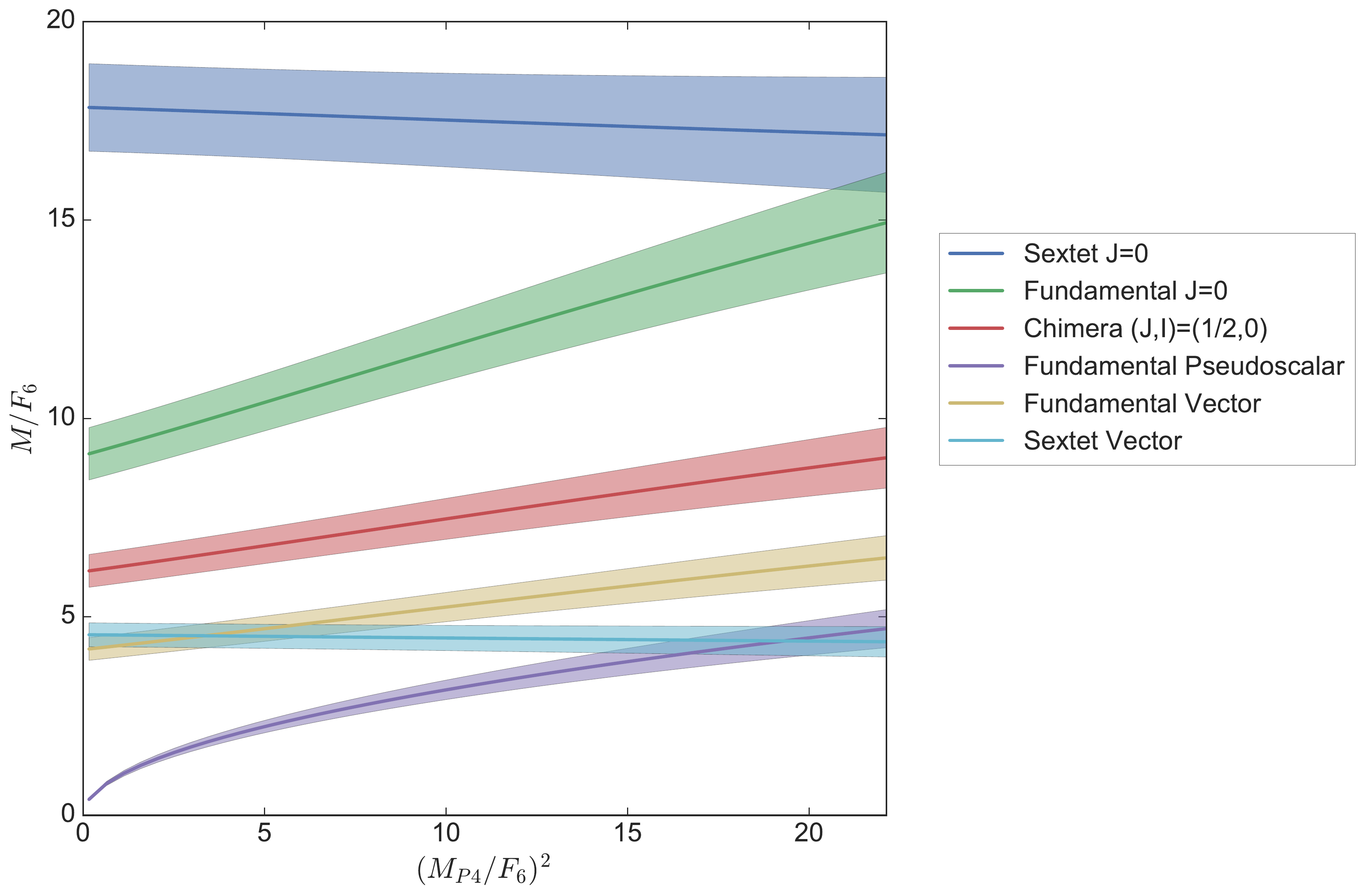}
\caption{
	Baryon and meson masses in the $m_6 \rightarrow 0$ limit.
        The chimera $(J,I)=(1/2,0)$ state corresponds to the top partner of Ferretti's model.
	The small rise of the sextet quantities in this limit is due to the mild variation of $\hat{F}_6$ with the fundamental fermion mass.
	Mesonic quantities were determined in~\cite{Ayyar:2017qdf}.
	\label{fig:pheno_overview_ferretti_limit}
}
\end{figure}

Table~\ref{table:dual_chiral_limit} gives the spectrum of light hadrons in our system in units of $F_6$, and Fig.~\ref{fig:pheno_overview_ferretti_limit} shows the baryon and meson masses in the $m_6 \rightarrow 0$ limit as a function of the ratio $(M_{P4}/F_6)^2$.
The mass ratio of $Qqq$ baryons to vector mesons is quite similar to what is seen in QCD.
The ratios of all masses to $F_6$ are smaller than in QCD, but that is something we have seen before, and is broadly consistent with large-$N_c$ expectations.

Current experimental evidence suggests that $\xi \ltapprox 0.1$ \cite{Bellazzini:2014yua,Panico:2015jxa,Aad:2015pla}, which means that the scale of the new strong sector is roughly $F_6 \gtapprox \sqrt{2} v/\sqrt{0.1} \simeq 1.1~\tev$ in our normalization.
Figure~\ref{fig:pheno_overview_ferretti_limit} then shows that the mass of the $\Lambda$ analogue---the top partner in Ferretti's model---must be $M_\Lambda \gtapprox 6.5~\tev$.
This estimate for the mass of the top partner will be modified by perturbative corrections from interactions with the Standard Model.
We expect these corrections to be small, just as perturbative electromagnetic corrections to hadron masses are small in QCD\@.
We note that the present work has not attempted a detailed budgeting of systematic effects from the lattice computation itself.
This includes, of course, those due to the slightly different fermion content of the model we studied in comparison with the  Ferretti model.

Although our results for the chimera mass indicate that it is somewhat heavier than assumed in Ref.~\cite{Ferretti:2014qta}, it remains to be seen whether this leads to any significant phenomenological tension or fine-tuning requirement.  The most crucial role of the top partner is in the generation of a realistic potential for the Higgs boson; we plan to investigate the top contribution to the Higgs potential non-perturbatively in a future work.  We are also planning a follow-up study of the decay matrix elements of the chimera baryon, which will allow the calculation of its decay width; experimental bounds on the top-partner mass typically assume a narrow width, and could be significantly weaker for a wide resonance.

\begin{acknowledgments}

Computations for this work were carried out with resources provided by the USQCD Collaboration, which is funded
by the Office of Science of the U.S.\ Department of Energy;  with the Summit supercomputer, a joint effort of the University of Colorado Boulder and Colorado State University, which is supported by the National Science Foundation (awards ACI-1532235 and ACI-1532236), the University of Colorado Boulder, and Colorado State University; by the Cy-Tera Project, which is co-funded by the European Regional Development Fund and the Republic of Cyprus through the Research Promotion Foundation; and the DECI resource ARCHER based in the United Kingdom at the University of Edinburgh with support from PRACE.

This work was supported in part by the U.S.\ Department of Energy under grant DE-SC0010005 (Colorado) and by the Israel Science Foundation under grants 449/13 and~491/17 (Tel Aviv).  Brookhaven National Laboratory is supported by the U. S. Department of Energy under contract DE-SC0012704.

\end{acknowledgments}

\appendix

% Appendices go here:
%\clearpage
\section{Data tables \label{app:data_tables}}
\begin{table}[h] 
\centering
\setlength{\tabcolsep}{10pt} % "Wider" columns
%\begin{tabular}{clllllc}
%\toprule
%Ensemble	&  $\beta$ & $\kappa_4$ & $ \kappa_6$ & $N_s$ &  $N_t$ & Configurations \\
%\hline
%1  &  7.25 &  0.13095 &  0.13418 &  16 &  32 &             61 \\
%2  &  7.25 &  0.13147 &  0.13395 &  16 &  32 &             71 \\
%3  &  7.30 &  0.13117 &  0.13363 &  16 &  32 &             61 \\
%4  &  7.30 &  0.13162 &  0.13340 &  16 &  32 &             71 \\
%5  &  7.55 &  0.13000 &  0.13250 &  16 &  32 &             84 \\
%6  &  7.65 &  0.12900 &  0.13080 &  16 &  32 &             49 \\
%7  &  7.65 &  0.13000 &  0.13100 &  16 &  32 &             84 \\
%8  &  7.65 &  0.13000 &  0.13200 &  16 &  32 &             84 \\
%9  &  7.75 &  0.12800 &  0.13100 &  16 &  32 &             84 \\
%10 &  7.75 &  0.12900 &  0.13080 &  16 &  32 &             54 \\
%11 &  7.75 &  0.12950 &  0.13150 &  16 &  32 &             34 \\
%12 &  7.85 &  0.12900 &  0.13080 &  16 &  32 &             44 \\
%\toprule
%\end{tabular}
\begin{tabular}{clllc}
\toprule
Ensemble	&  $\beta$ & $\kappa_4$ & $ \kappa_6$  & Configurations \\
\hline
1  &  7.25 &  0.13095 &  0.13418 &  61 \\
2  &  7.25 &  0.13147 &  0.13395 &  71 \\
3  &  7.30 &  0.13117 &  0.13363 &  61 \\
4  &  7.30 &  0.13162 &  0.13340 &  71 \\
5  &  7.55 &  0.13000 &  0.13250 &  84 \\
6  &  7.65 &  0.12900 &  0.13080 &  49 \\
7  &  7.65 &  0.13000 &  0.13100 &  84 \\
8  &  7.65 &  0.13000 &  0.13200 &  84 \\
9  &  7.75 &  0.12800 &  0.13100 &  84 \\
10 &  7.75 &  0.12900 &  0.13080 & 54 \\
11 &  7.75 &  0.12950 &  0.13150 & 34 \\
12 &  7.85 &  0.12900 &  0.13080 & 44 \\
\toprule
\end{tabular}
\caption{
	The ensembles list used in this study.
	All ensembles have volume $V = N_s^3 \times N_t = 16^3 \times 32$.
}
\label{table:ensembles}
\end{table}

\begin{table}[h] 
\centering
\setlength{\tabcolsep}{10pt} % "Wider" columns
\begin{tabular}{clll}
\toprule
Ensemble	 & $t_0/a^2$ & $\hat{m}_4$ & $\hat{m}_6$ \\
\hline
1  &  1.093(9) &  0.0422(7) &   0.020(1) \\
2  &  1.135(9) &   0.028(1) &   0.025(1) \\
3  &   1.13(1) &  0.0345(8) &   0.032(1) \\
4  &  1.111(9) &  0.0228(6) &  0.0381(8) \\
5  &   1.85(2) &   0.050(1) &   0.034(1) \\
6  &  1.068(5) &   0.082(1) &  0.0896(8) \\
7  &   1.46(2) &   0.046(2) &   0.080(2) \\
8  &   2.29(2) &   0.038(1) &   0.036(2) \\
9  &   1.56(1) &   0.108(1) &   0.071(1) \\
10 &   1.75(2) &   0.073(2) &   0.077(2) \\
11 &   2.62(2) &   0.047(1) &   0.040(1) \\
12 &   2.67(2) &   0.060(1) &   0.060(1) \\
% two-digit uncertainties
%1  &  1.0925(90) &  0.04222(70) &   0.0203(10) \\
%2  &  1.1351(91) &   0.0279(11) &   0.0251(12) \\
%3  &   1.132(12) &  0.03454(83) &   0.0323(14) \\
%4  &  1.1113(89) &  0.02284(63) &  0.03814(76) \\
%5  &   1.845(18) &   0.0495(11) &   0.0340(13) \\
%6  &  1.0675(52) &   0.0816(10) &  0.08959(83) \\
%7  &   1.463(15) &   0.0459(18) &   0.0801(22) \\
%8  &   2.294(22) &   0.0382(13) &   0.0357(21) \\
%9  &   1.556(12) &   0.1077(12) &   0.0708(10) \\
%10 &   1.754(15) &   0.0730(19) &   0.0771(16) \\
%11 &   2.621(20) &   0.0465(13) &   0.0402(14) \\
%12 &   2.670(22) &   0.0602(14) &   0.0599(12) \\
\toprule
\end{tabular}

\caption{
	Fermion masses and flow scales for the ensembles used in this study.
}
\label{table:fermion_masses}
\end{table}

\begin{table}[h] 
\centering
\setlength{\tabcolsep}{12pt} % "Wider" columns
\begin{tabular}{clll}
\toprule
Ensemble	 & Chimera $(J,I)=(1/2,0)$ & Chimera $(J,I)=(1/2,1)$ & Chimera $(J,I)=(3/2,1)$ \\
\hline
1  &             0.84(3) &             0.80(3) &             0.94(6) \\
2  &             0.80(3) &             0.75(3) &             0.91(6) \\
3  &             0.84(2) &             0.81(2) &             0.95(4) \\
4  &             0.83(3) &             0.80(3) &             0.96(6) \\
5  &             0.97(3) &            0.960(9) &             1.13(6) \\
6  &             1.11(2) &             1.09(1) &             1.20(2) \\
7  &             1.04(2) &             1.02(2) &             1.15(3) \\
8  &             1.00(4) &             0.96(4) &             1.15(4) \\
9  &             1.24(3) &             1.21(2) &             1.34(5) \\
10 &             1.17(2) &             1.14(3) &             1.28(3) \\
11 &             1.07(4) &             1.03(4) &             1.19(5) \\
12 &             1.17(2) &             1.13(3) &             1.30(3) \\
% two-digit uncertainties
%1  &             0.835(26) &             0.799(29) &             0.935(57) \\
%2  &             0.798(34) &             0.750(34) &             0.910(55) \\
%3  &             0.839(20) &             0.808(18) &             0.949(42) \\
%4  &             0.825(33) &             0.798(31) &             0.957(64) \\
%5  &             0.969(27) &            0.9595(85) &             1.131(57) \\
%6  &             1.112(16) &             1.089(13) &             1.201(16) \\
%7  &             1.039(20) &             1.019(20) &             1.153(34) \\
%8  &             1.003(35) &             0.960(35) &             1.147(42) \\
%9  &             1.238(32) &             1.211(20) &             1.340(49) \\
%10 &             1.169(20) &             1.143(25) &             1.283(30) \\
%11 &             1.071(43) &             1.034(38) &             1.185(59) \\
%12 &             1.168(22) &             1.134(33) &             1.296(32) \\
\toprule
\end{tabular}
\caption{
	Masses $\hat{M}_{Qqq}$	for the chimera baryons in units of the flow scale $t_0/a^2$.
}
\label{table:chimera_baryons}
\end{table}

\begin{table}[h] 
\centering
\setlength{\tabcolsep}{12pt} % "Wider" columns
\begin{tabular}{clll}
\toprule
Ensemble & Fundamental $(J=0)$ & Fundamental $(J=1)$ & Fundamental $(J=2)$ \\
\hline
1  &       1.13(7) &       1.20(8) &       1.30(9) \\
2  &       1.07(9) &        1.1(1) &       1.26(9) \\
3  &       1.10(7) &       1.13(8) &       1.25(8) \\
4  &        1.0(1) &        1.1(1) &        1.3(3) \\
5  &       1.33(4) &       1.39(4) &       1.54(8) \\
6  &       1.46(3) &       1.50(4) &       1.61(5) \\
7  &       1.29(8) &       1.37(5) &        1.5(2) \\
8  &       1.36(6) &        1.4(1) &        1.6(2) \\
9  &       1.75(2) &       1.79(4) &       1.85(7) \\
10 &       1.54(5) &       1.60(3) &       1.72(6) \\
11 &        1.5(2) &       1.55(9) &        1.7(2) \\
12 &       1.53(4) &       1.61(3) &       1.73(6) \\
% Two-digit uncertainties
%1  &       1.134(74) &       1.195(82) &       1.304(85) \\
%2  &       1.070(92) &        1.13(12) &       1.255(93) \\
%3  &       1.101(73) &       1.132(77) &       1.254(81) \\
%4  &        1.05(14) &        1.10(11) &        1.26(25) \\
%5  &       1.327(43) &       1.394(35) &       1.537(77) \\
%6  &       1.464(27) &       1.504(36) &       1.605(48) \\
%7  &       1.293(83) &       1.368(50) &        1.49(18) \\
%8  &       1.358(61) &        1.44(10) &        1.59(15) \\
%9  &       1.750(22) &       1.787(39) &       1.852(65) \\
%10 &       1.538(51) &       1.597(28) &       1.719(64) \\
%11 &        1.48(18) &       1.551(94) &        1.65(16) \\
%12 &       1.531(39) &       1.611(26) &       1.728(55) \\
\toprule
\end{tabular}
\caption{
	Masses $\hat{M}_{q^4}$ for the fundamental baryons in units of the flow scale $t_0/a^2$.
}
\label{table:fundamental_baryons}
\end{table}

\begin{table}[h] 
\centering
\setlength{\tabcolsep}{12pt} % "Wider" columns
\begin{tabular}{cllll}
\toprule
Ensemble & Sextet $(J=0)$ & Sextet $(J=1)$ & Sextet $(J=2)$ & Sextet $(J=3)$ \\
\hline
1  &  1.89(7) &  1.92(7) &  2.00(6) &  2.13(6) \\
2  &  1.90(6) &  1.95(7) &   2.1(1) &   2.2(1) \\
3  &  1.880(3) &  1.93(4) &  2.05(6) &  2.17(5) \\
4  &  1.98(5) &  2.04(5) &  2.11(5) &  2.23(6) \\
5  &  2.27(2) &  2.33(3) &  2.46(2) &  2.61(2) \\
6  &   2.6(2) &  2.63(8) &   2.7(2) &  2.77(7) \\
7  &  2.76(4) &  2.81(5) &  2.90(5) &  3.00(4) \\
8  &  2.49(4) &  2.54(6) &  2.67(7) &  2.84(4) \\
9  &  2.55(8) &  2.57(7) &  2.66(6) &  2.80(4) \\
10 &  2.75(6) &  2.79(6) &   2.9(2) &  3.01(5) \\
11 &  2.54(4) &  2.58(6) &  2.68(6) &  2.81(7) \\
12 &  2.71(5) &  2.74(4) &  2.89(4) &  3.06(5) \\
% Two-digit uncertainty
%1  &  1.892(73) &  1.915(66) &  1.999(59) &  2.132(56) \\
%2  &  1.903(58) &  1.948(67) &   2.05(13) &   2.18(14) \\
%3  &  1.878(32) &  1.933(43) &  2.045(56) &  2.170(45) \\
%4  &  1.982(46) &  2.035(52) &  2.107(53) &  2.227(63) \\
%5  &  2.273(24) &  2.330(25) &  2.461(24) &  2.607(24) \\
%6  &   2.59(17) &  2.632(79) &   2.70(17) &  2.767(65) \\
%7  &  2.761(41) &  2.809(45) &  2.895(47) &  3.004(44) \\
%8  &  2.486(44) &  2.539(62) &  2.666(74) &  2.841(41) \\
%9  &  2.554(78) &  2.565(74) &  2.656(59) &  2.797(38) \\
%10 &  2.745(64) &  2.790(55) &   2.89(17) &  3.011(50) \\
%11 &  2.536(40) &  2.577(58) &  2.681(58) &  2.813(65) \\
%12 &  2.707(45) &  2.739(44) &  2.889(41) &  3.056(52) \\
\toprule
\end{tabular}
\caption{
	Masses $\hat{M}_{Q^6}$ for the sextet baryons in units of the flow scale $t_0/a^2$.
}
\label{table:sextet_baryons}
\end{table}

\section{Technical matters---lattice \label{app:lattice}}
\subsection{Chimera baryons}

Let $Q$ denote a sextet fermion and $q$ a fundamental fermion.
The interpolating field for a chimera baryon has the form $\mathcal{O}_B^\epsilon = \epsilon_{abcd} Q_\alpha^{ab} q_\gamma^c q_\delta^d C^{\alpha\gamma\delta\epsilon}$, where Latin indices indicate SU(4) color and Greek indices indicate spin.
For brevity we suppress flavor SU(2) indices.
The tensor $C$ is some combination of gamma matrices which projects onto the desired spin state.
Because these chimera operators are fermionic, they naturally carry a free spinor index.
We find it useful to work in a ``non-relativistic" formulation, projecting onto eigenstates of $P_{\pm} = \frac{1}{2}(1 \pm \gamma_4)$.
This projection produces two-component spinors, which we identify with the familiar spin-up and spin-down states of a non-relativistic fermion.
To extract the ground-state mass from a two-point correlation function, any gauge-invariant operator with the correct spin and internal quantum numbers suffices.
Since the mass spectrum is the focus of the present work, we find the non-relativistic formulation easiest to implement.
For a discussion of its use in the existing lattice literature, see~\cite{Liu:1998um,Leinweber:2004it} and references therein.

Propagators form the numerical building blocks of our correlation functions:
\begin{align}
D^{-1}_q(m|n)^{a,b}_{\alpha,\beta} \equiv \left\langle q(m)^a_{\alpha} \bar{q}(n)^b_\beta \right\rangle,
\end{align}
where $m,n$ are points on the lattice; $a,b$ are $\SU(4)$-color indices; and $\alpha,\beta$ are non-relativistic spin indices.
There is an analogous expression for the sextet propagator $D^{-1}_Q$.
A chimera propagator then takes the form
\begin{align}
	\begin{split}
		\left\langle \mathcal{O}^\lambda_B(m) \overline{\mathcal{O}}^\zeta_B(n) \right\rangle
		&= \epsilon_{abcd} \epsilon_{efgh} C^{\alpha\gamma\delta\lambda} C^{\epsilon\phi\eta\zeta} D^{-1}_Q(m|n)^{ab,ef}_{\alpha,\epsilon} \\
		& \phantom{XXXX}\times \left[ D^{-1}_q(m|n)^{c,g}_{\gamma,\phi} D^{-1}_q(m|n)^{d,h}_{\delta,h} - D^{-1}_q(m|n)^{c,h}_{\gamma,\eta} D^{-1}_q(m|n)^{d,g}_{\delta,\phi} \right] \label{eq:chimera_2pt}
	 \end{split}
\end{align}
The bracketed expression contains both a direct and an exchange term.
Both terms are necessary for states like the charged $\Sigma$ or $\Sigma^*$ in QCD which consist of a single light flavor $u$ or $d$.
States like the $\Lambda$, $\Sigma_0$, or $\Sigma_0^*$ inherently contain light quarks of two different flavors $u$ and $d$.
Since different valence flavors cannot be contracted, such states possess no exchange term.
For the chimera analogues of the $\Sigma$ and $\Sigma^*$, we consider $I_z = 1$ states, which include both the direct and exchange term.

The spin projectors $C^{\alpha\beta\gamma\lambda}$ isolate the correct spin states for the initial and final baryons.
%They may be regarded as encoding Clebsch-Gordan coefficients which couple the spins, isolating the correct spin states for the initial and final baryons.
For example, the standard decomposition of the spins of the $S=-1$ hyperons is:
\begin{align}
&\Sigma^*: \ket{J=3/2,I=1} = \ket{\up \up \up} \notag \\
&\Sigma: \ket{J=1/2,I=1} = \frac{1}{\sqrt{6}} \left[ 2 \ket{\dn \up \up} - \ket{\up \up \dn} - \ket{\up \dn \up} \right] \notag \\
&\Lambda: \ket{J=1/2,I=0} = \frac{1}{\sqrt{2}} \left[ \ket{\up \up \dn} - \ket{\up \dn \up} \right].
\end{align}
In each line we have taken the state with largest value of $J_z$: for example, \hbox{$\ket{J=1/2,I=0}$} is shorthand for the $J_z = + 1/2$ state.
The states on the {right-hand side are $\ket{ S_z^Q S_z^q S_z^q}$.

\subsection{Spectroscopy}

For baryon two-point correlation functions, we use a smeared (Gaussian) source operator on the $t=0$ time slice and a point operator at the sink, projecting onto zero spatial momentum.
Smearing is done after fixing to the Coulomb gauge.
In order to achieve strong signals and flat effective masses, we used smearing radii ranging from $r_0 = 4a$ to $12a$.
We use anti-periodic boundary conditions in the temporal direction for the fermion propagators.
In order to maximize statistics, we combine the correlation functions for the forward-propagating and backward-propagating eigenstates of $P_+$ and $P_-$.
After tuning the smearing radius $r_0$ to achieve flat effective masses, we model the baryon two-point functions using a single decaying exponential.
Our final fitting procedure for baryons follows that described in~\cite{Ayyar:2017qdf}.
In particular, all baryon results quoted in the present work include systematic uncertainty from the choice of $[t_\text{min}, t_\text{max}]$.

Details relating to fits for mesonic quantities and fermion masses are described in~\cite{Ayyar:2017qdf}.

\subsection{Global fit details}

Figure~\ref{fig:fit_pulls} shows the distribution of pulls from our global fit of the baryon masses to Eqs.~(\ref{eq:sextet_mass_model}), (\ref{eq:fundamental_mass_model}), and~(\ref{eq:chimera_mass_model}).
Pulls---roughly, the difference between the fit and the data in units of the error of the difference---provide a straightforward and useful test for bias in large fits~\cite{Demortier:Pulls}.
In the asymptotic limit where the number of data points becomes large, the pull distribution should approach a unit-width $(\sigma=1)$ normal distribution centered at the origin $(\mu=0)$.

\begin{figure}[t]
	\includegraphics[width=0.5\textwidth]{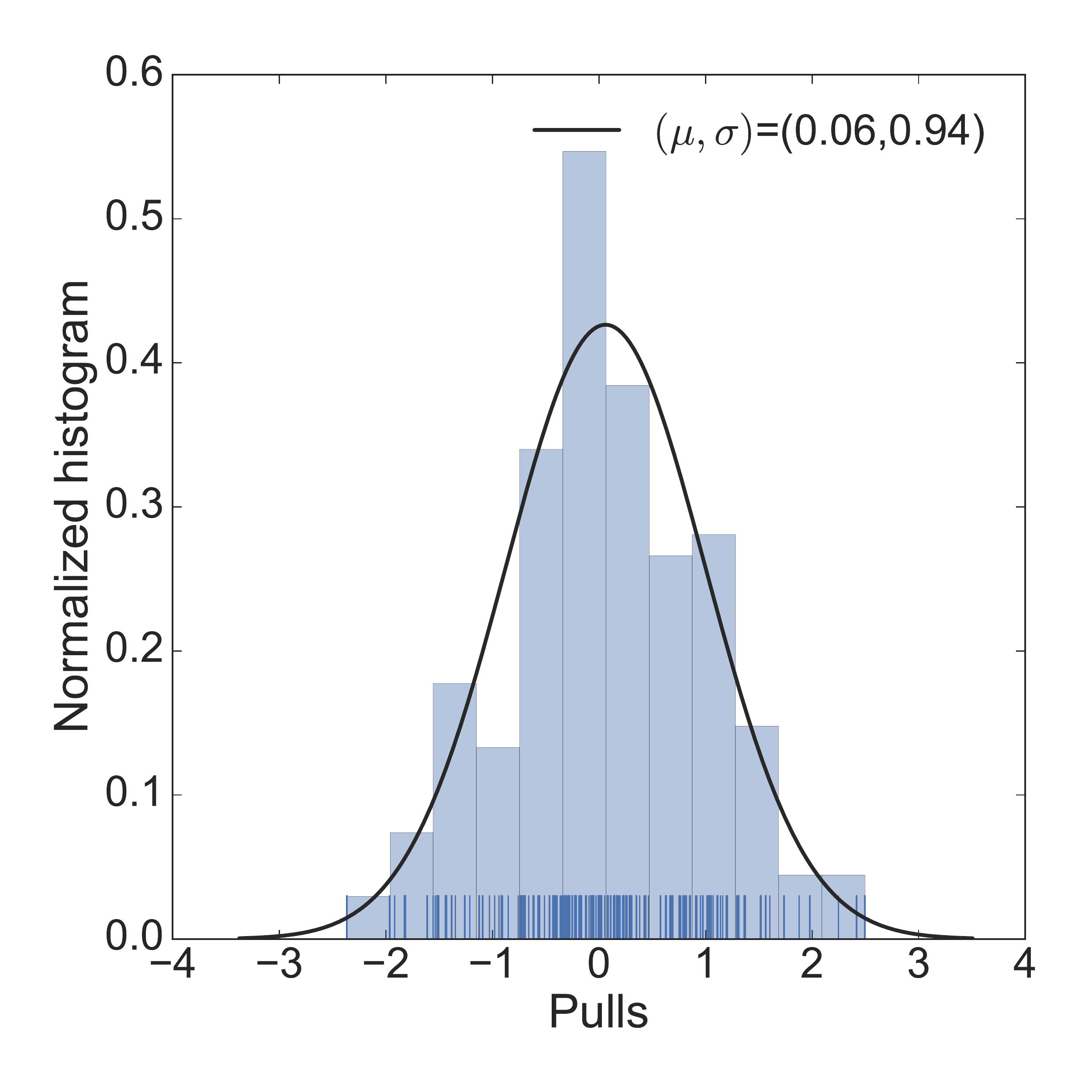}
	\caption{
		The distribution of pulls from the fit.
		\label{fig:fit_pulls}
	}
\end{figure}

\bibliography{su4_baryon_spectro}

\end{document}